\documentclass[prb,aps,longbibliography,floats,superscriptaddress,twocolumn]{revtex4-2}

\usepackage[utf8]{inputenc}
\usepackage{amsmath,amssymb,amsthm,amsbsy,bm}
\usepackage{graphicx,epsfig,epstopdf,subfigure}
\usepackage{dsfont,verbatim,varwidth,dcolumn}
\usepackage[usenames,dvipsnames]{xcolor}
\usepackage{color,soul,subfiles}
\usepackage[colorlinks,hypertexnames=false]{hyperref}

\makeatother

\begin{document}

\title{Parafermions in a multilegged geometry: Towards a scalable parafermionic network}
\author{Udit Khanna}
\altaffiliation{Present address: Department of Physics, Bar-Ilan University, Ramat Gan 52900, Israel}
\email{udit.khanna.10@gmail.com}
\affiliation{Raymond and Beverly Sackler School of Physics and Astronomy, Tel Aviv University, Tel Aviv 6997801, Israel}
\affiliation{Department of Condensed Matter Physics, Weizmann Institute of Science, Rehovot 76100, Israel}
\author{Moshe Goldstein}
\affiliation{Raymond and Beverly Sackler School of Physics and Astronomy, Tel Aviv University, Tel Aviv 6997801, Israel}
\author{Yuval Gefen}
\affiliation{Department of Condensed Matter Physics, Weizmann Institute of Science, Rehovot 76100, Israel}

\begin{abstract}
  Parafermionic zero modes are non-Abelian excitations which have been predicted to emerge at  
  the boundary of topological phases of matter. Contrary to earlier proposals, here we show that such zero 
  modes may also exist in multilegged star junctions of quantum Hall states. 
  We demonstrate that the quantum states spanning the degenerate parafermionic Hilbert space may be detected and manipulated 
  through protocols employing quantum antidots and fractional edge modes.
  Such star-shaped setups may be the building blocks of two-dimensional parafermionic networks.
\end{abstract}

\maketitle

\textit{Introduction. }
Parafermion (PF) zero modes are fractionalized excitations with non-Abelian statistics, which may exist
at the boundary of certain topological phases of matter~\cite{AliceaReview}.
PF zero modes are expected to show a number of interesting phenomena~\cite{Alicea1,Stern1,Luiz2017,Stern2,Fibonacci2,Alicea2,App1,App2,App3,
Zhang1,App4,App6,App7,App8,App9,App10,App11,App12,Tiwari17,Jelena2021}
such as the fractional Josephson effect, zero-bias anomaly, and topological Kondo effect, 
and may be potentially useful in quantum information applications~\cite{Stern2,NayakRMP,QIC1,QIC2,QIC3,QIC4,QIC5,QIC6,QIC7,QIC8}.
General classifications of truly one-dimensional (1D) bosonic and fermionic systems rule out the existence of PFs beyond 
Majorana zero modes~\cite{Class1,Class2,Class3,Class4}. However, PF zero modes may 
be supported at suitably designed line junctions at the boundary of fractional quantum Hall (QH) 
states~\cite{Alicea1,Stern1,Luiz2017}. In particular, two counterpropagating chiral edge modes with opposite spins, at 
the interface of fractional quantum Hall (FQH) puddles in the $\nu = 1/m$ state, may be gapped by proximity to a bulk superconductor or a 
ferromagnet. The domain walls between superconducting and ferromagnetic segments are expected to host 
$\mathbb{Z}_{2m}$ PF zero modes on the boundary~\cite{Alicea1,Stern1}. 
Similar setups have been proposed for the $\nu = 2/3$ state~\cite{Stern2}, in hierarchical~\cite{Hierarchy} and 
bilayer~\cite{Bilayer1,Bilayer2,App5} fractional QH phases, in proximitized and fractional 
topological insulators~\cite{Zhang1,Loss2,Tiwari15,Loss15,GlazmanPRL,Tiwari17,Jelena19,Trauzettel19}, 
as well as in systems of coupled 1D wires~\cite{Fendley12,Sela14,Loss1,Loss14,Manisha17,Mross18,Mora18,Ronetti21}.
Such PF modes can be detected through their transport signatures in appropriately designed setups~\cite{Alicea2,App5,App12,App13,Kyrylo1,Jelena2021}.
Two-dimensional parafermionic networks are predicted to host even more exotic Fibonacci anyons, which may eventually
allow fault-tolerant universal quantum computation~\cite{Fibonacci1,Fibonacci2,Fibonacci3,Stern2}. 

\begin{figure}[t]
  \centering
  \includegraphics[scale=0.23]{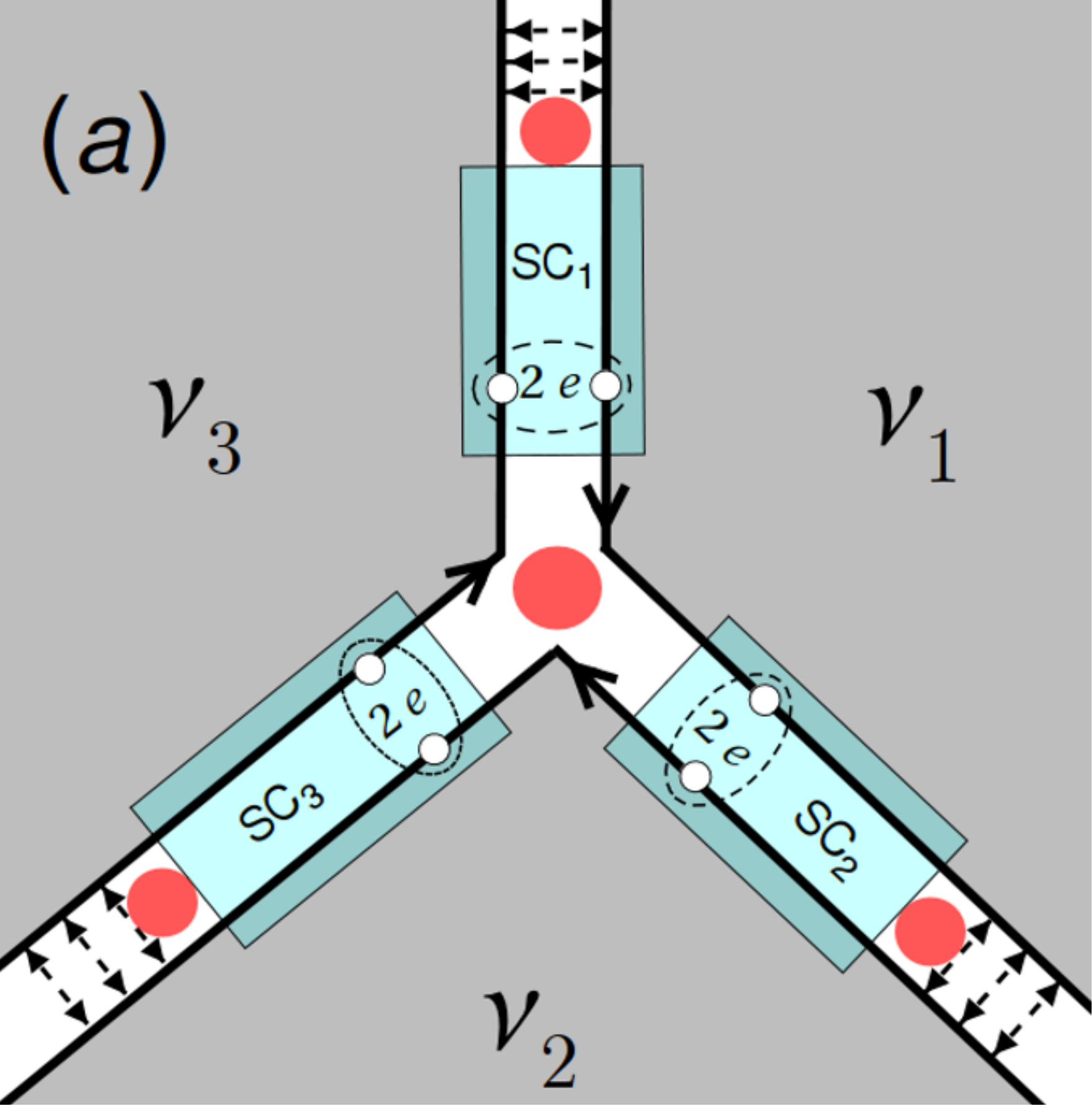}
  \includegraphics[scale=0.23]{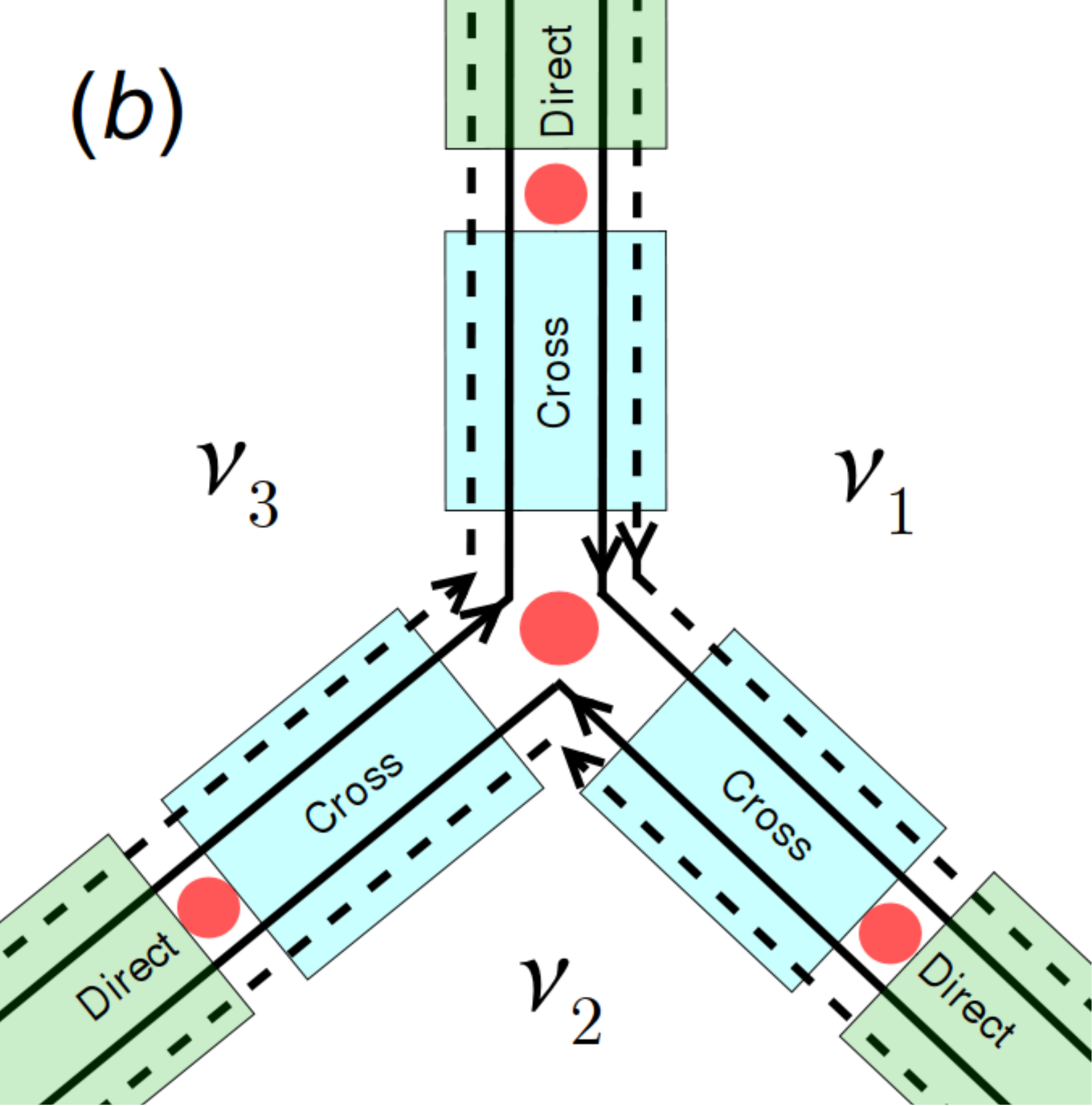}
  \caption{ Trijunctions hosting parafermionic zero modes (represented 
  by red circles) may be constructed in (a) single-layer or 
  (b) double-layer setups. The black lines represent the chiral edge
  modes of the three quantum Hall puddles. In (b), the solid (dashed) lines
  represent the edge modes of the top (bottom) layer. 
  Counterpropagating pairs of edge modes may be gapped 
  out by introducing superconducting or electron-tunneling segments in the
  single-layer setup [panel (a)]. In the bilayer setup [panel (b)], edge modes
  may be gapped out by introducing segments with direct (intralayer) or crossed (interlayer)
  electron tunneling. }
\end{figure}

Here we propose a setup hosting PF zero modes, based on multilegged star junctions of FQH states.
Figure~1 shows the simplest such geometry supporting a parafermionic zero mode localized at the center of the star junction. 
Such a geometry may be based on single-layer FQH states~\cite{Alicea1,Stern1} employing superconducting regions [Fig.~1(a)] or 
double-layer states~\cite{Bilayer1,Bilayer2} employing interlayer electron tunneling at the interfaces [Fig.~1(b)]. 
Similar setups have been studied previously in the context of Majorana zero modes (which are $\mathbb{Z}_{2}$ PFs) 
localized at the boundaries of 1D nanowires~\cite{TJ0,Beenakker1,TJ3,TJ1,TJ2,TJ4,TJ5,TJ6,TJ7}. 
These studies found that a single zero-energy Majorana mode is supported at the center of odd-legged star junctions, but not
in the case of an even number of legs. The present work is a concrete extension of the previous studies to the case of $m > 1$ 
PF modes, which requires one to consider interacting systems. Here, one might naively expect a modulo $2m$ dependence on the 
number of wires. In contrast, we interestingly find the modulo 2 behavior to persist, namely, that parafermionic junctions 
with an odd (even) number of legs would (would not) host a single parafermion at the center
of the star which cannot (can) be gapped out.
Below we analyze the low-energy dynamics of a trijunction geometry, demonstrating the existence of parafermionic zero mode 
and the degeneracy of the zero-energy Hilbert space. Furthermore, we propose specific setups employing quantum antidots 
(QADs) and fractional edge modes, which may be used to measure the degeneracy of the parafermionic Hilbert space 
and facilitate manipulation of such states.

\textit{Trijunction model. }
We consider a single trijunction setup [Fig.~1(a)] comprising the boundary of three $\nu_{j} = \nu = \frac{1}{m}$ 
($j = 1, 2, 3$) QH puddles with same spin polarization. The low-energy dynamics of each puddle 
is governed by a chiral edge mode, described by a bosonic field $\phi_{j} (x, t)$~\cite{Wen1990}. These bosonic fields 
are described by the Hamiltonian 
\begin{align} \label{eq:1md}
  H_{\text{edge}} &= \frac{m v}{4 \pi} \sum_{j = 1}^{3} \int dx\, \big[ \partial_x \phi_{j} (x) \big]^2, 
\end{align}
where $v$ is the edge mode velocity (assumed to be identical, for all puddles, for simplicity) and satisfy
\begin{align} \label{eq:2md}
  [\phi_j (x), \phi_j (y)] &= \frac{i \pi}{m} \text{sgn}(x - y) , \\
  [\phi_{j} (x), \phi_{k} (y)] &= i \frac{\pi}{m} \text{ for all } j > k . \label{eq:3md}
\end{align}
Here the coordinate $x$ increases along the direction of propagation for each edge mode. 
The second commutator above relies on the assumption that the edge modes are segments of a single boundary. 
At the center of the trijunction, each pair of counterpropagating edge modes is gapped out through 
proximity coupling to a $p$-wave superconductor~\cite{Foot1}. The superconducting regions are described by 
\begin{align} \label{eq:4md}
  H_{s} &= \Delta \sum_{j} \int dx \, \cos \big[ m \big(\phi_{j} + \phi_{j-1} \big) + \hat{\varphi} \big],  
\end{align}
where $\phi_{0}$ has been identified with $\phi_{3}$, and we assume the amplitude ($\Delta$) and phase 
($\hat{\varphi}$) of all superconducting segments to be identical. 
This is possible if all three segments are generated by proximity to the same bulk superconductor. 
In order to fix the boundary conditions and the total charge hosted by the setup, we 
also assume that (away from the junction) each pair of counterpropagating modes is gapped out through strong 
interedge electron tunneling [denoted as black dashed double-headed arrows in Fig.~1(a)]. 
The tunneling regions are described by
\begin{align} \label{eq:5md}
  H_{t} &= g \sum_{j} \int dx \, \cos \big[ m \big( \phi_{j} - \phi_{j-1} \big) \big].
\end{align}
Finally, we assume that the superconducting segments (implicitly assumed to be part of the same bulk superconductor) 
are floating. Their total charge (${Q}_{T}$) appears in the Hamiltonian as a charging term, $(\hat{Q} - {Q}_{T})^{2} / 2C$,  
where $\hat{Q} = \sum_{j} \hat{Q}_{j}$ is the total charge and $\hat{Q}_j$ is 
the charge hosted in the $j{\text{th}}$ superconducting segment, and is given by
\begin{align}
  \hat{Q}_{j} &= \frac{1}{2\pi} \int_{\text{SC}_{j}} dx \, \big( \partial_{x} \phi_{j} - \partial_x \phi_{j-1} \big).
\end{align}
Since the (total) charge and phase of the superconductor are conjugate variables, they satisfy 
$[\hat{\varphi}, \hat{Q}] = 2i$. 

We restrict our analysis to energies below $\sqrt{4 \pi m v \Delta}$ and $\sqrt{4 \pi m v g}$. In this limit, the
fractional chirals may be assumed to be completely pinned inside the superconducting and tunneling segments~\cite{Stern1}. 
It follows that the superconducting and tunneling regions are described by integer-valued operators 
$\hat{N}_{j}$ and $\hat{M}_{j}$, respectively. Therefore at each boundary of the trijunction, we must have
\begin{align}
  \phi_{j} + \phi_{j-1} = \frac{2\pi}{m} \hat{N}_j , 
\end{align}
where $\hat{N}_{j} \rightarrow \hat{N}_{j} + \hat{\varphi}/m$. 
Similarly, at each boundary of the tunneling regions we have
\begin{align}
  \phi_{j} - \phi_{j-1} &= \frac{2\pi}{m} \hat{M}_{j} .
\end{align}

Our next task is to analyze the low-lying dynamics of this trijunction block showing that it exhibits parafermionic
physics. To this end, we perform a standard mode expansion of the chiral fields, 
\begin{align} \nonumber
  \phi_{j} (x, t) = \varphi_j + &A_j p_{\varphi_{j}} \big( v t - x \big) \\ + \sum_{n > 0} 
    &\bigg[ B_{n j} a_{n j} e^{-i q_{n j} (v t - x )} + \text{H.c.} \bigg],
\end{align}
where $A_{j}$ and $B_{n j}$ are $c$ numbers and $a_{n j}$ describe the bosonic excitations in the chiral modes. 
Imposing the boundary conditions [Eq.~(7)] and commutation relation [Eq.~(2)] inside the trijunction, we find
\begin{align}
  \phi_1 (x,t) &= \varphi_{1} + \sqrt{\frac{2}{m}} \sum_{n > 0} \bigg[ \frac{a_{n}}{\sqrt{2n-1}}
  e^{-i q_{n} (vt - x)} + \text{H.c.} \bigg], \\
  \phi_2 (x,t) &= -\phi_1 (x + L_1,t) + \varphi_{1} + \varphi_{2}, \\
  \phi_3 (x,t) &= -\phi_2 (x + L_2,t) + \varphi_{2} + \varphi_{3},
\end{align}
where $\varphi_{j} = \frac{\pi}{m} \big( \hat{N}_{j} + \hat{N}_{j+1} - \hat{N}_{j-1} \big)$
(not to be confused with $\hat{\varphi}$), 
$q_{n} = (2n-1)\frac{\pi}{\ell} $, $L_{j}$ is the length of the $j{\text{th}}$ chiral inside
the trijunction, and $\ell = \sum_{j} L_{j}$.   
A similar expansion can be performed for the region between the superconductor and the tunneling segments~\cite{Alicea1}.

Plugging the expansion found above into Eq.~(3), we find that the only nonzero commutation relations are 
\begin{align}
  [\hat{N}_3, \hat{N}_2 ] &= \frac{i m}{\pi} \, \,\,\,\,\, \text{ and} \\
  [\hat{N}_1, \hat{M}_1 ] &= [\hat{N}_2, \hat{M}_1 ] = [\hat{N}_3, \hat{M}_1 ] = \frac{i m}{\pi}.
\end{align}
Using (6), the charge on the superconducting segments (at low energies) is found to be
\begin{align}
  \hat{Q}_{j} &= \frac{1}{m} (\hat{N}_{j+1} - \hat{N}_{j-1} - \hat{M}_{j}). 
\end{align}
Note that the total charge $\hat{Q} = -\sum_{j} \hat{M}_{j}$ depends only on the integers describing the 
electron-tunneling regions at the boundaries. 
Since the charge on each superconducting segment is defined module 2, the physical operators to consider are
$ e^{i \pi \hat{Q}_j} $. These  satisfy
\begin{align}
  e^{i \pi \hat{Q}_j} e^{i \pi \hat{Q}} &= e^{i \pi \hat{Q}} e^{i \pi \hat{Q}_j} \,\, \text{ and} \\
  e^{i \pi \hat{Q}_{j-1}} e^{i \pi \hat{Q}_{j}} &= e^{i \pi/ m} e^{i \pi \hat{Q}_{j}} e^{i \pi \hat{Q}_{j-1}} 
  \,\, \text{ for all } j. 
\end{align}
Finally, we may write the Hamiltonian of the trijunction (ignoring the charging term for now) as 
\begin{align}
  H = H_{\text{edge}} + H_{s} + H_{t} = \sum_{n > 0} v q_{n} \big(a_{n}^{\dagger} a_{n} + \frac{1}{2}\big). 
\end{align}
Note that $H$ only depends on the bosonic excitations of the chiral fields and not on $\hat{N}$. 
This allows for the possibility that the Hamiltonian supports zero-energy solutions. To this end, we construct
operators which commute with the Hamiltonian and describe (local) superpositions of quasiholes and quasiparticles
in the chiral edge modes. A quick calculation confirms that the operator $\tilde{\Gamma}$ defined as 
\begin{align} \label{eq:19md}
  \tilde{\Gamma}_{0} &= e^{i \varphi_{1}} \int dx \, \sum_{k = 1}^{3} 
  \big[ e^{-i \varphi_{k}} e^{i \phi_k (x)} + \text{ H.c.} \big] 
\end{align}
satisfies these requirements~\cite{SM}. We may define $\Gamma_{0} = e^{i \varphi_{1}} = e^{i \pi/ m  (\hat{N}_1 + \hat{N}_2 - \hat{N}_3 )}$ 
as the projection of $\tilde{\Gamma}_{0}$ onto the ground state (in the absence of bosonic fluctuations). 
Similar expressions can be defined for the three zero modes ($\Gamma_{j}$) localized at the domain wall between 
the ($j{\text{th}}$) superconducting segment and corresponding tunnel junction at the boundary of the trijunction. 
Therefore the low-energy physics of the trijunction is governed by the four localized zero-energy modes which satisfy 
parafermionic commutation relations, 
$\Gamma_{\alpha} \Gamma_{\beta} = e^{i \pi/ m } \Gamma_{\beta} \Gamma_{\alpha}$ if $\alpha$ appears to the left
of $\beta$ in the tuple $(1, 0, 2, 3)$. 
We note that there are multiple ways to represent the PF zero modes, which are localized at
the different boundaries of the junction (for instance, $\Gamma_{0}$ as defined above is localized at 
$\phi_{1}$). Here we have used the convention that $\Gamma_{j}$ ($j = 1, 2, 3$) is localized at $\phi_{j}$. 

The low-energy subspace of the trijunction is expected to be degenerate due to the PF commutation relations of 
the zero modes. Since the operators $ e^{i \pi \hat{Q}_j} $ commute with $H$, the different states 
in the ground state sector may be labeled through their eigenvalues. The non-trivial
commutations among these operators imply that the Hilbert space is spanned by states of the form 
$| Q_{\text{T}}, Q_{j} \rangle $ (for a fixed $j$). Since $e^{i \pi \hat{Q}_j}$ can 
assume $2m$ values, the total degeneracy of the ground state of the trijunction is $(2m)^2$, which is consistent with 
the presence of four $\mathbb{Z}_{2m}$ PF modes. Note that the energy of states with different $Q_{T}$ is set 
externally through the charging term in the Hamiltonian [which was ignored in eq.~(18)]. Therefore the degeneracy of 
such states is lifted for finite charging energy. By contrast, the $2m$ states with a given $Q_{T}$ are degenerate 
up to exponentially small splittings (arising from the finite length of the superconductor)~\cite{Burnell16} which are neglected here. 

Our analysis of the trijunction may be extended to star-shaped setups with a larger number of legs. The results are
qualitatively unaltered for stars with an odd number of legs. On the other hand, for even-legged junctions, the 
low-energy Hamiltonian of the junction [$H$ in Eq.~(18)] involves additional terms which explicitly depend 
on the integer-valued operators $\hat{N}_{j}$. Such terms rule out the possibility of a $\mathbb{Z}_{2m}$ 
PF zero mode at the center of an even-legged junctions. Our results are also applicable to star-shaped junctions 
in bilayer setups. Specifically, a trijunction comprising the edge of the QH bilayer [Fig.~1(b)] with $\nu = 1/m$ 
per layer would host a $\mathbb{Z}_{m}$ PF zero mode at its center. 

\textit{Ground state manipulation. }
Here we discuss protocols (adopted from Ref.~\cite{Kyrylo1}) to induce transitions between the degenerate low-energy 
states of the trijunction. As described above, the total charge of the trijunction ($Q_{T}$) is fixed externally through the charging energy.  
However, the degenerate states within a topological sector (labeled by $Q_{T}$), may be manipulated through 
redistribution of the charge among the superconducting segments.
Suppose $|\Psi\rangle$ is an eigenstate of $ e^{i \pi \hat{Q}_1} $ with eigenvalue $e^{i \pi Q_{1}} = 1$; 
Eqs.~(15) and (16) then imply that $ \big( e^{i \pi \hat{Q}_2} \big)^{k} |\Psi\rangle$
is also an eigenstate of $ e^{i \pi \hat{Q}_1} $ with eigenvalue $e^{i (\pi/m) k }$. Therefore one may induce transitions 
between $|Q_{T}, Q_{1} \rangle$ and $ |Q_{T}, Q_{1}+\frac{1}{m} \rangle $ through the application of 
$ e^{i \pi \hat{Q}_{2}} $. Similarly, application of $ e^{i \pi \hat{Q}_{3}} $ induces transitions from 
$|Q_{T}, Q_{1} \rangle$ to $ |Q_{T}, Q_{1}-\frac{1}{m} \rangle $. 

\begin{figure*}[t]
  \centering
  \includegraphics[scale=0.25]{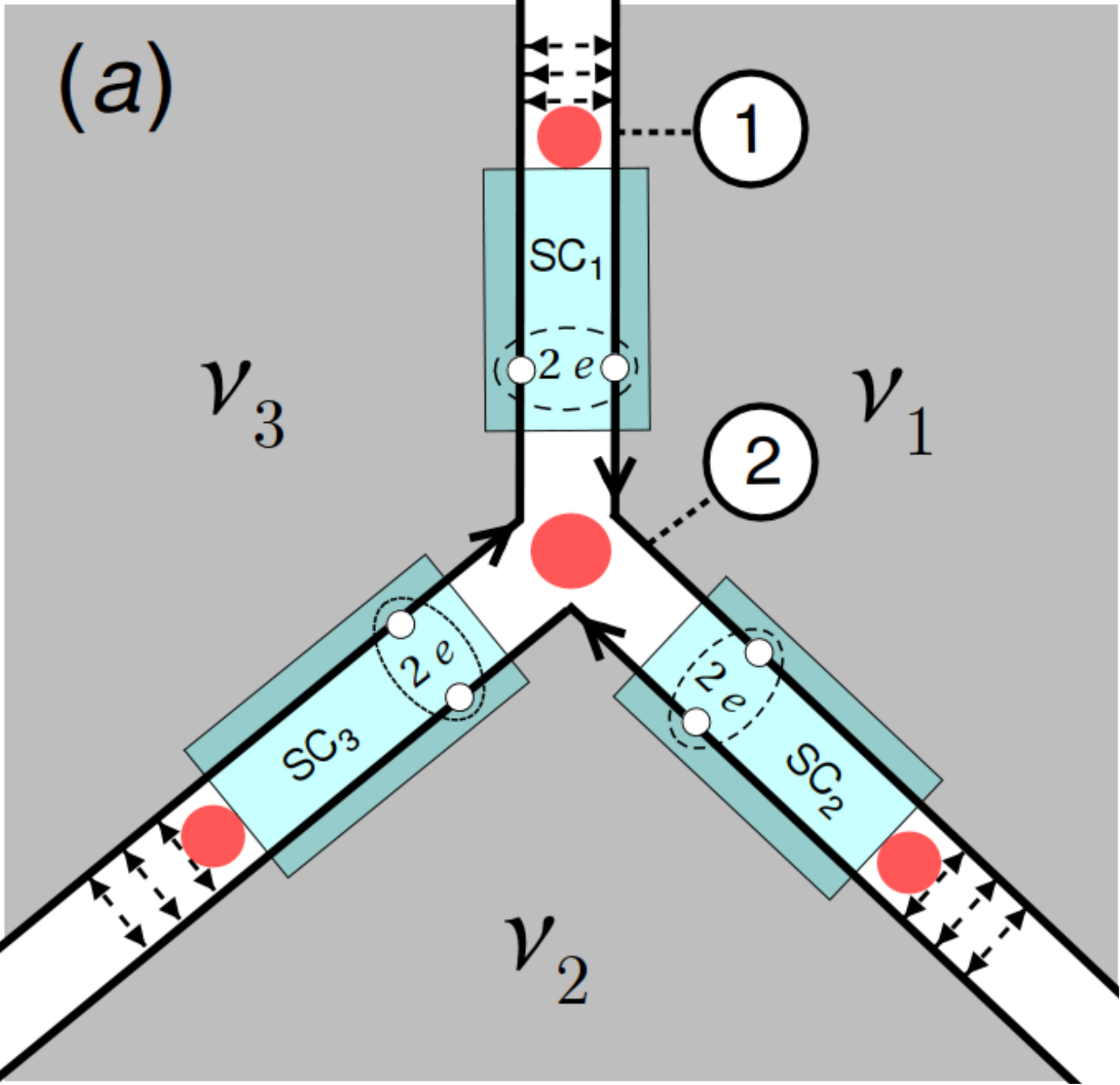}
  \includegraphics[scale=0.25]{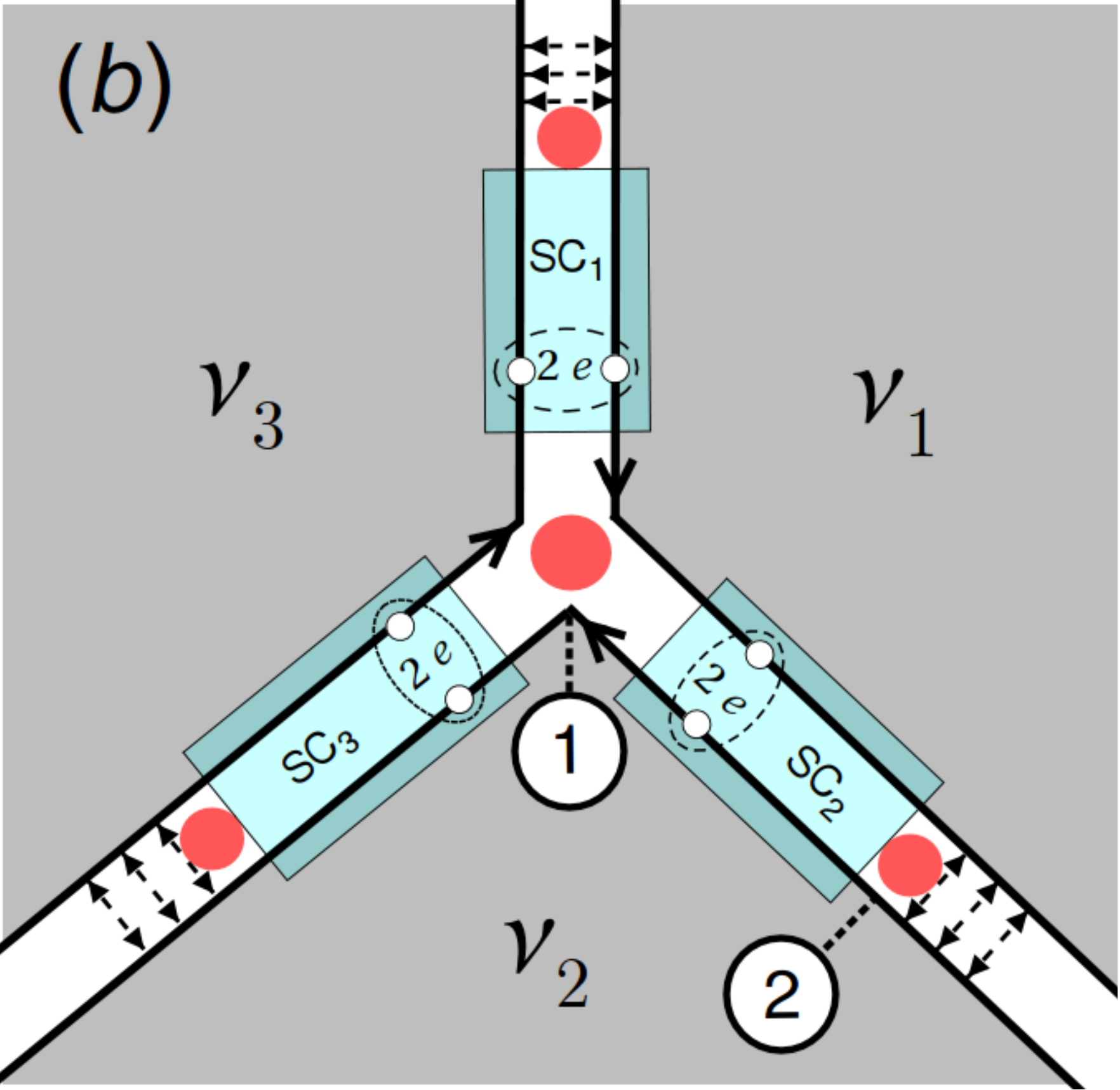}
  \includegraphics[scale=0.25]{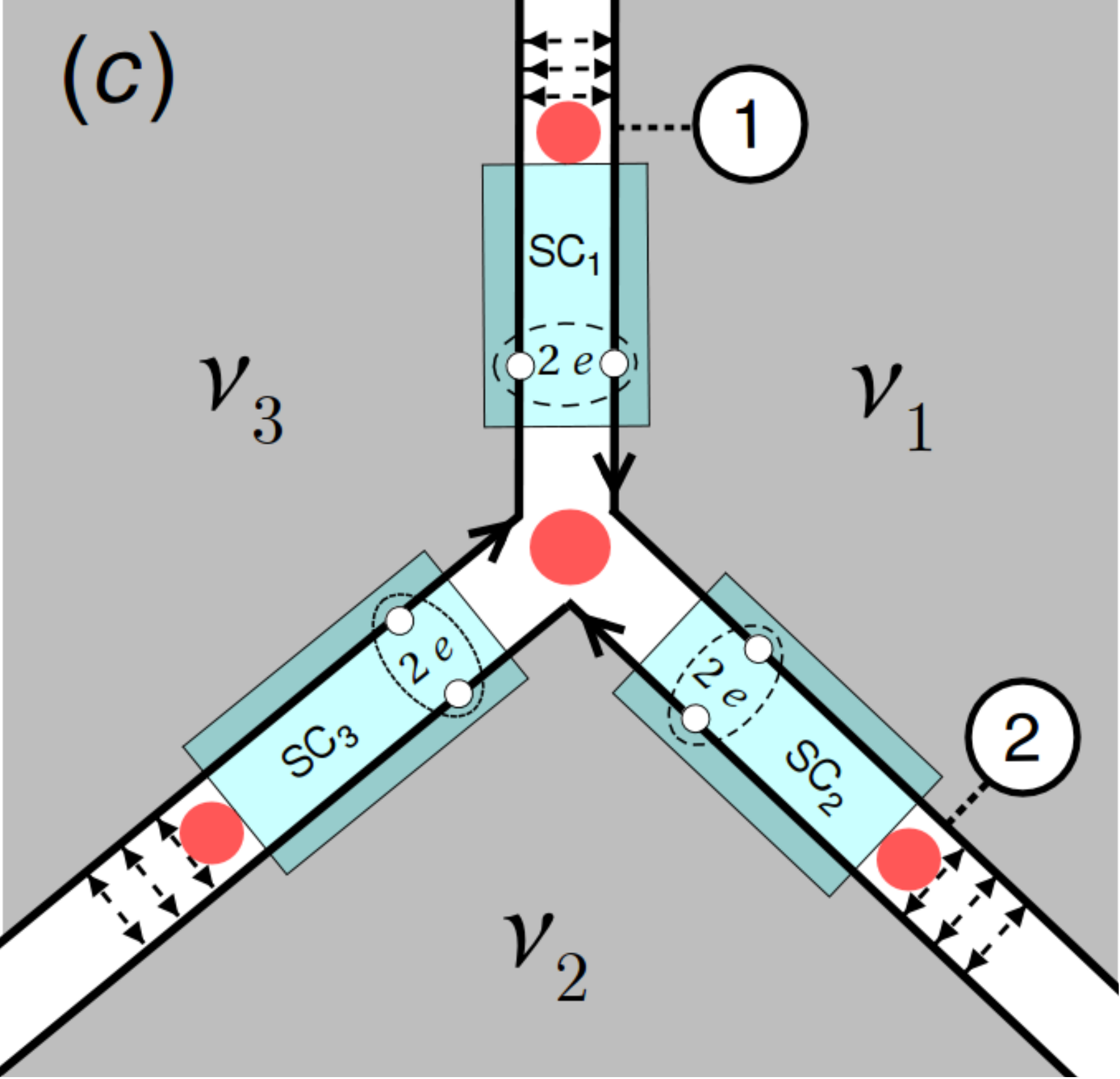}
  \caption{ Schemes for state manipulation in the trijunction (adopted from Ref.~\cite{Kyrylo1}). The white circles denote quantum 
  antidots which may host a single Laughlin quasiparticle and are tunnel coupled to one of the interfaces supporting 
  a parafermionic mode. As described in the text, the state of the trijunction may be manipulated through 
  the application of operators of the form $e^{\pm i \pi \hat{Q}_{j}}$ ($j = 1, 2, 3$). Varying the gate voltage
  on the antidots in an adiabatic fashion allows for the application of 
  (a) $e^{\pm i \pi \hat{Q}_1}$, (b) $e^{\pm i \pi \hat{Q}_2}$, and (c) $e^{\pm i \pi (\hat{Q}_1 + \hat{Q}_2)} \sim 
  e^{\pm i \pi (\hat{Q} - \hat{Q}_{3})}$. 
  Moving the antidots to the opposite side of the same interface would modify the operator by an overall phase.}
\end{figure*}

Figure~2 depicts several setups involving two QADs coupled to the trijunction, which may be 
used to apply the operators $e^{i \pi \hat{Q}_{j}}$. We assume that each QAD may be empty or host a single Laughlin quasiparticle 
depending on the voltage applied. Thus effectively each QAD can be modeled by a two-level system with the Hamiltonian,
\begin{align}
  H_{\text{QAD}} = V_{Q} \bigg( N_{\psi} - \frac{1}{2} \bigg), 
\end{align}
where $V_{Q}$ is proportional to the electrostatic potential on the QAD and $N_{\psi}$ is the number of quasiparticles in 
the QAD. The tunnel coupling of the QADs to the PF modes in the trijunction may be described by 
\begin{align}
  H_{\text{tun}} = \sum_{k = 1,2} \tilde{J}_k \psi_k \Gamma_{k}^{\dagger} + \text{H.c.}, 
\end{align}
where $\psi_k$ is the quasiparticle operator for the $k{\text{th}}$ QAD satisfying $[N_{\psi_{k}}, \psi_{k}] = -\psi_{k}$, 
and $\Gamma_k$ is the PF operator coupled to the QAD. Since
the PF modes have multiple representations (which are localized at different edges), here
$\Gamma_{k}$ is assumed to be localized at the boundary of the QH region in which the QAD is located. 
We assume that the tunneling amplitudes ($\tilde{J}_k$)
are much smaller than the charging energy of the trijunction ($E_c$) so that $H_{\text{tun}}$ leads to cotunneling of
quasiparticles between the QADs, which may be described by 
\begin{align}
  H_{\text{co-tun}} = J \psi_{1} \psi_{2}^{\dagger} \Gamma_{1}^{\dagger} \Gamma_{2} + \text{H.c.} 
\end{align}
Here $J \sim \tilde{J}_{1} \tilde{J}_{2}^{*} / E_{c}$ for $E_{c} \gg \tilde{J}_{1,2}$. 
Let us assume that the QADs were initially decoupled from 
the trijunction and prepared such that one QAD is occupied while the other is empty. Then by coupling them to the trijunction 
and slowly varying the voltage on the dots, we may induce the transition 
$|\Psi\rangle \rightarrow  \Gamma_{1}^{\dagger} \Gamma_{2} | \Psi \rangle$.
Using Eqs. (13)--(15), we find that the terms of the form $\Gamma_{0} \Gamma_{j}^{\dagger}$ are proportional 
to the operators $e^{i \pi \hat{Q}_j}$. We may facilitate the application of any operator ($e^{i \pi \hat{Q}_{j}}$)
through suitable placement of the two QADs. 
The setups shown in Figs.~2(a)--(c) may be used to realize $e^{i \pi \hat{Q}_{1}}$, $e^{i \pi \hat{Q}_2}$, and $e^{i \pi \hat{Q}_3}$, 
respectively. 

\begin{figure}[b]
  \centering
  \includegraphics[scale=0.25]{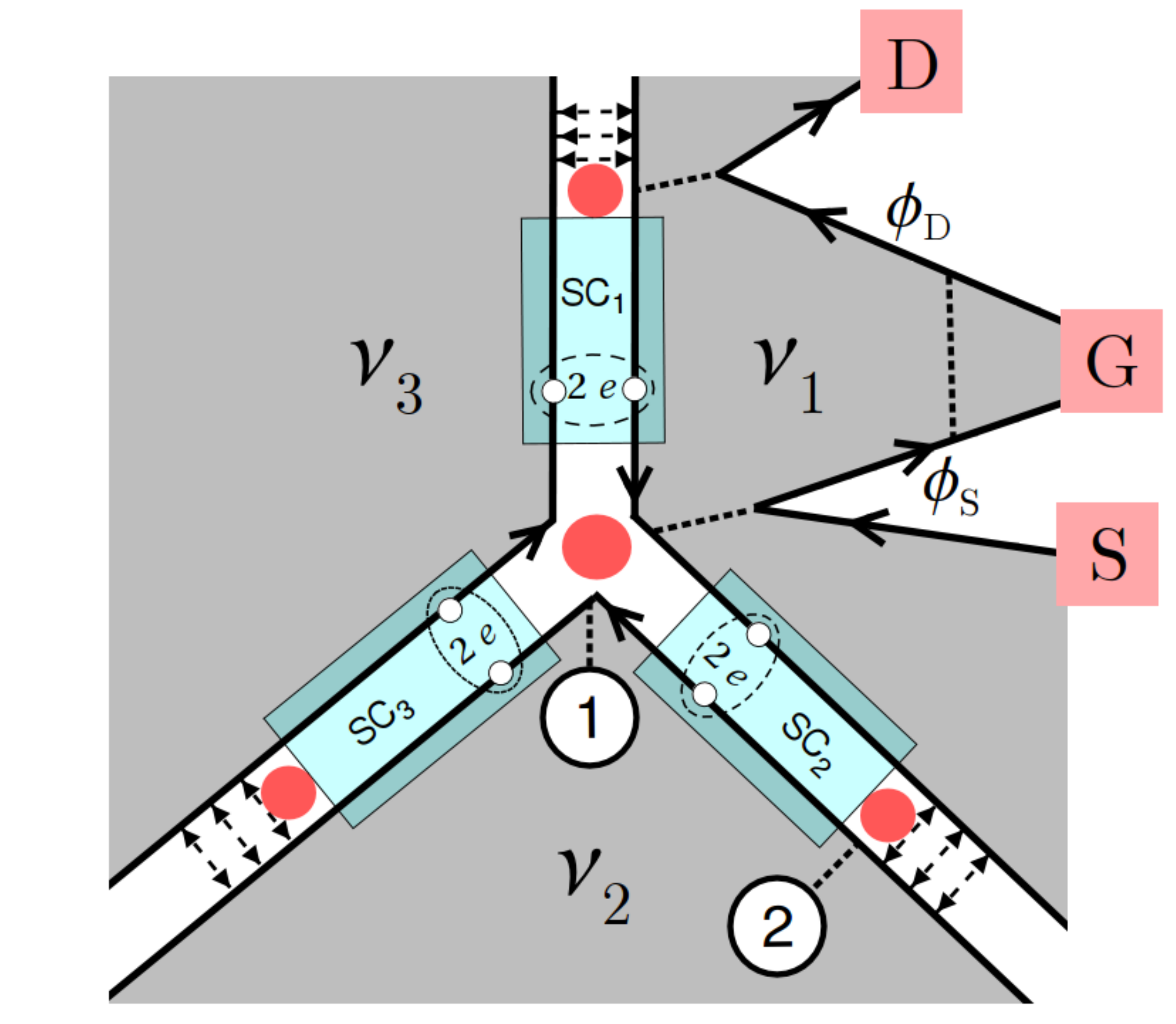}
  \caption{ Scheme for measuring the degeneracy of the trijunction (adopted from Ref.~\cite{Kyrylo1}). The white circles denote
  quantum antidots which facilitate the application of $e^{\pm i \pi \hat{Q}_2}$ upon the 
  trijunction state. $\phi_{S/D}$ are additional fractional edge modes tunnel coupled to the trijunction
  and to each other. Quasiparticles emanated from the source ($S$) may end up at the drain ($D$),
  by directly tunneling from $\phi_{S}$ to $\phi_{D}$, or by tunneling across the trijunction. 
  The interference between these two processes, and therefore the current ($I_{t}$) from $S$ to $D$, is 
  sensitive to the state of the trijunction. Therefore repeatedly applying $e^{i \pi \hat{Q}_{2}}$ through 
  the antidots and monitoring $I_{t}$, allows one to measure the number of degenerate states in the parafermionic
  Hilbert space of the trijunction (for a fixed total charge).}
\end{figure}

\textit{Degeneracy of parafermionic states. }
For a given total charge ($Q_{T}$), the degeneracy of the low-energy Hilbert space of the trijunction is 
expected to be $2m$. Here we propose a setup (depicted in Fig.~3), involving QADs and two external fractional
QH edge modes, to directly observe the degenerate subspace. The external fractional modes act as leads from 
which fractional quasiparticles may tunnel into the trijunction. 
The Hamiltonian of the two edge modes is given by $H_{a} = (mv/4\pi) \int dx [ \partial_x \phi_{a} ]^2
- V_a \int dx  \partial_x \phi_{a}$, where $V_a$ is the voltage on $\phi_{a}$ ($a = s, d$).
The quasiparticle tunneling from the edges to the trijunction is described by 
\begin{align}
  H_{\text{qp}} = \sum_{a = s,d} \tilde{\eta}_a e^{i \phi_a} \Gamma_{a}^{\dagger} + \text{H.c.}
\end{align}
As pointed out above, the representation of the
PF mode has to be appropriately chosen so that fractional charge only tunnels across the bulk of
a QH state, and not across vacuum. 
If the charging energy for the trijunction is large, then only cotunneling processes are allowed at low energies. 
These are described by   
$H_{\text{ct}} = \eta_{c} \Gamma_{1} \Gamma_{0}^{\dagger} e^{-i \phi_{d}} e^{i \phi_{s}} + \text{H.c.}$, 
where $\Gamma_0$ is the PF located inside the trijunction and $\Gamma_{1}$ is the
PF located at a domain wall between a superconductor and tunneling segment (cf. Fig.~3). 
We assume that quasiparticles may also tunnel between the two modes directly (without involving the trijunction). 
Such processes are described by
$H_{\text{dir}} = \eta_{d} e^{-i \phi_{d}} e^{i \phi_{s}} + \text{H.c.}$
Then the total tunneling Hamiltonian is $H_{\text{dir}} + H_{\text{ct}} = (\eta_{d} + \eta_{c} \Gamma_{1} \Gamma_{0}^{\dagger})
e^{-i \phi_{d}} e^{i \phi_{s}} + \text{H.c.}$, which induces a finite
 current from the source S to the drain D in Fig.~3.
The tunneling current may be evaluated in the limit of weak tunneling, following the analysis of Wen~\cite{Wen91PRB}, as
\begin{align}
  I_{t} = \frac{|\eta_{\text{eff}}|^2}{v^{2 \nu}} \frac{2 \pi \nu}{\Gamma(2 \nu)} (\nu |V_{1} - V_{2}|)^{2\nu - 1}, 
\end{align}
where $\eta_{\text{eff}}$ is the eigenvalue of $\eta_{d} + \eta_{c} \Gamma_{1} \Gamma_{0}^{\dagger}$.   
As shown earlier, the operator $\Gamma_{j} \Gamma_{0}^{\dagger}$ depends on $e^{i \pi \hat{Q}_{j}}$ 
($e^{i \pi \hat{Q}_{1}}$ for the setup in Fig.~3). This implies that the current $I_{t}$  
between the edge modes is sensitive to the state of the trijunction, which may be labeled using $e^{i \pi \hat{Q}_{1}}$. 
In fact, the state of the trijunction would be projected to an eigenstate of $e^{i \pi \hat{Q}_{1}}$ upon 
measurement of the tunneling current ($I_{t}$).

The QADs in the setup shown in Fig.~3 may be used to apply the operator $e^{i \pi \hat{Q}_{2}}$ to the state of the 
trijunction, which induces a transtion between the degenerate states labeled by $e^{i \pi Q_{1}}$. As discussed above, 
the tunneling current $I_{t}$ is expected to change after the application of $e^{i \pi \hat{Q}_{2}}$, as it depends on 
the state of the trijunction.  
Therefore, the degeneracy of the trijunction (for a given charge $Q_{T}$)
may be measured by repeatedly monitoring $I_{t}$ and applying $e^{i \pi \hat{Q}_{2}}$
through the QADs. The number of steps required before $I_{t}$ returns to its initial value is the number of 
degenerate states in the PF Hilbert space. 

\begin{figure}[t]
  \centering
  \includegraphics[scale=0.22]{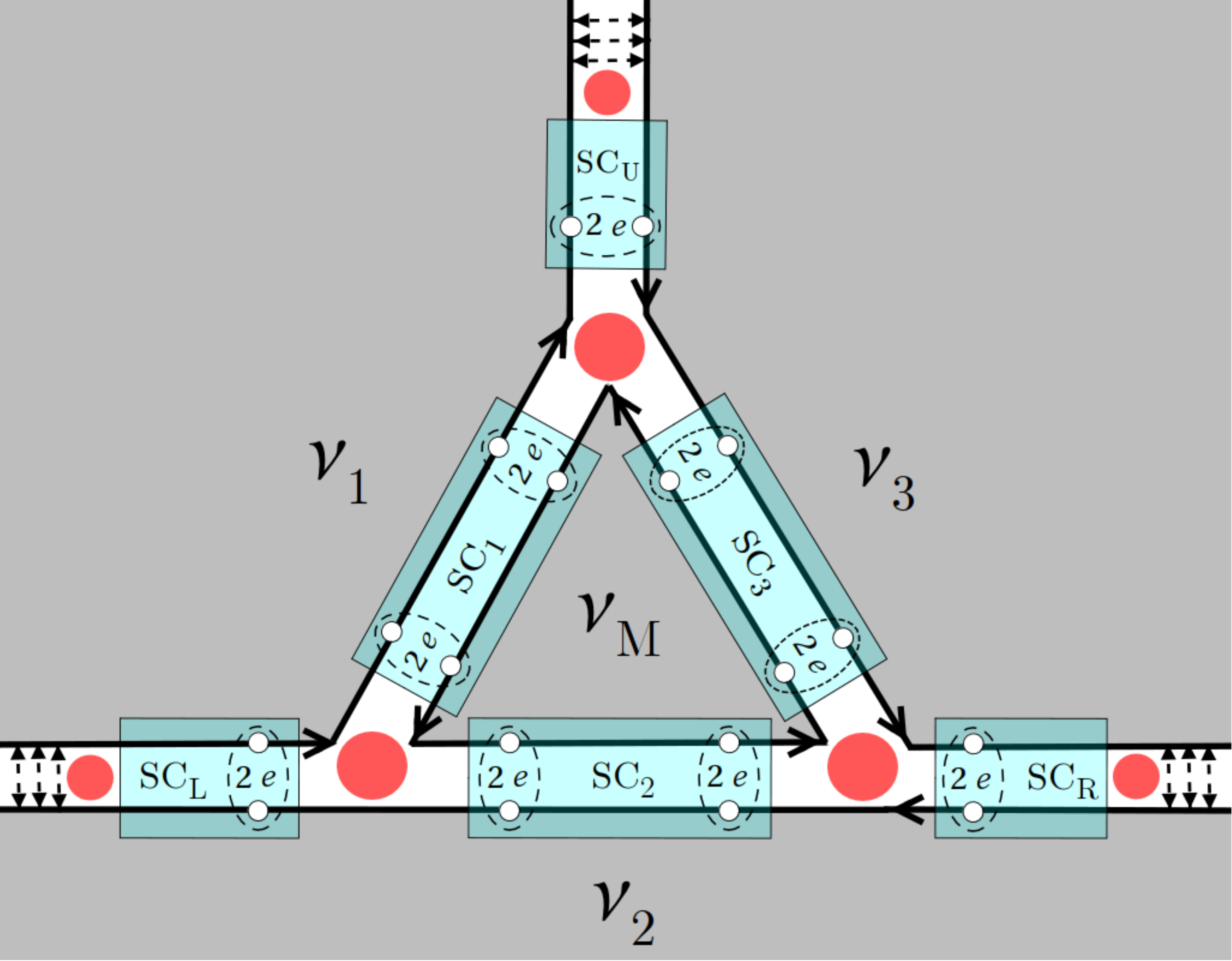}
  \caption{ Multiple trijunctions may be connected to form complex parafermionic networks. 
  The figure depicts a combination of three trijunctions, with a quantum Hall island 
  enclosed only by parafermionic zero modes. }
\end{figure}

\textit{Prospects of parafermionic networks. }
The PF zero modes, hosted at the center of star junctions, are identical to those realized in 
line junction setups~\cite{Alicea1,Stern1}. However, a crucial advantage of  our proposal is that 
such junctions may be used as building blocks for constructing complex parafermionic networks. 
We first note that the connectivity of a single trijunction, depicted in Fig.~1, is higher than 
the case of a four PF line junction. Such a configuration could be useful for implementing complex operations 
(such as, braiding~\cite{QIC2}) on the PF modes. 
Going on to a small number of star junctions may be combined in a myriad of configurations, each of which may 
be useful for easily implementing a different set of operations. 
Figure~4 depicts one such configuration involving three trijunctions, which involves a QH island surrounded 
only by PF zero modes. Repeating the mode expansion analysis for such {\it loop} geometries, we find 
that the PF states are sensitive to the fractional charge in the bulk of the enclosed QH puddle, which 
in turn may be manipulated through additional QADs. Configurations involving several such loops could be 
useful for quantum information applications. Finally, several such junctions may be employed for constructing 
large PF networks with different topologies, such as a cycle-free Bethe lattice or a two-dimensional 
honeycomb structure. The braiding properties of PFs moving on such networks depend sensitively on 
their connectivity~\cite{Tomasz0,Tomasz1,Tomasz2}. Additionally at low energies, such networks may be described in terms of emergent 
phases that support even more exotic non-Abelian excitations~\cite{Stern2}. 

\textit{Conclusions. }
We have demonstrated that setups involving multilegged star junctions of fractional quantum Hall states, may 
serve as a platform which hosts parafermionic zero modes. Specifically, employing a low-energy mode expansion, we showed
that a star with odd (even) number of legs supports (does not support) a parafermionic mode at its center. 
We also proposed setups involving additional quantum antidots and fractional edge modes, which facilitate the 
detection and manipulation of the parafermionic states. The star-shaped junction proposed here, could be used 
as building blocks for constructing complex parafermionic networks, including two-dimensional lattices, which 
may be quite difficult to realize using just line junctions. A detailed investigation of the emergent physics 
of such networks is left for the future.

{\it Acknowledgments. }
We acknowledge useful discussions with Kyrylo Snizhko. 
U. K. was supported by the Raymond and Beverly Sackler Faculty of Exact Sciences at Tel Aviv University 
and by the Raymond and Beverly Sackler Center for Computational Molecular and Material Science. 
M.G. was supported by the US-Israel Binational Science Foundation 
(Grant No.~2016224). Y.G. was supported by CRC~183 (project~C01), the Minerva Foundation, 
DFG Grants No.~RO~2247/11-1 and No.~MI~658/10-1, 
the German Israeli Foundation (Grant No.~I-118-303.1-2018), the Helmholtz International Fellow Award, and 
by the Italia-Israel QUANTRA grant.

\onecolumngrid
\clearpage

\setcounter{affil}{0}
\setcounter{page}{1}
\renewcommand{\thefigure}{S\arabic{figure}}
\setcounter{figure}{0}
\renewcommand{\theequation}{S\arabic{equation}}
\setcounter{equation}{0}
\renewcommand\thesection{S\arabic{section}}
\setcounter{section}{0}

\title{Supplementary Material for ``Parafermions in a multilegged geometry: Towards a scalable parafermionic network''}
\author{Udit Khanna}
\altaffiliation{Present address: Department of Physics, Bar-Ilan University, Ramat Gan 52900, Israel}
\email{udit.khanna.10@gmail.com}
\affiliation{Raymond and Beverly Sackler School of Physics and Astronomy, Tel Aviv University, Tel Aviv 6997801, Israel}
\affiliation{Department of Condensed Matter Physics, Weizmann Institute of Science, Rehovot 76100, Israel}
\author{Moshe Goldstein}
\affiliation{Raymond and Beverly Sackler School of Physics and Astronomy, Tel Aviv University, Tel Aviv 6997801, Israel}
\author{Yuval Gefen}
\affiliation{Department of Condensed Matter Physics, Weizmann Institute of Science, Rehovot 76100, Israel}

\begin{abstract}

This supplemental material provides additional details regarding our analysis of the trijunction. 

\end{abstract}

\maketitle

\begin{centering}
\subsection*{Parafermions in the Trijunction}
\end{centering}

As described in the main text, the $j{\text{th}}$ superconducting segment ($j = 1,2,3$) 
imposes the following boundary condition on the chiral modes entering ($\phi_{j-1}$) and emanating ($\phi_{j}$) from it, 
\begin{align} \label{eq:bc}
  \phi_{j} (x = 0, t) + \phi_{j-1} (x = L_{j-1}, t) = \frac{2 \pi}{m} \hat{N}_{j}, 
\end{align}
where $j = 0$ has been identified with $j = 3$ for brevity. Additionally, we use the convention that $x$ increases along
the direction of propagation for each chiral mode. Note that physically, the boundary condition (\ref{eq:bc}) 
implies that the current entering the superconductor is equal in magnitude and opposite in sign to the current 
exiting the superconductor. To find the spectrum of the problem, we consider the standard mode expansion for right-moving 
chiral fields for the three chirals,
\begin{align} \nonumber
  \phi_{j} (x, t) = \varphi_j + &A_j p_{\varphi_{j}} \big( v t - x \big) + \\\sum_{n > 0}
    &\bigg[ B_{n j} a_{n j} e^{-i q_{n j} (v t - x )} + \text{h.c.} \bigg]. \label{eq:exp}
\end{align}
Here, $A_{j}$ and $B_{n j}$ are $c$-numbers, $\varphi_{j}$, $p_{\varphi_{j}}$ are the conjugate pair of zero-mode operators, 
and $a_{n j}$ is the annihilation operator for the $n^{\text{th}}$ ($n \geq 1$) bosonic excitation with wavevector $q_{n j}$. 
Using (\ref{eq:bc}) in (\ref{eq:exp}), we find, 
\begin{align} \nonumber
  \varphi_{j} + &A_{j} p_{\varphi_{j}} v t = \\ \label{eq:z1}
  &- \varphi_{j-1}  - A_{j-1} p_{\varphi_{j-1}} \big( v t - L_{j-1} \big) + \frac{2\pi}{m} \hat{N}_{j}, \\
  \sum_{n} B_{n j} &a_{n j} e^{-i q_{n j} v t} = \nonumber \\ \label{eq:z2} 
  &-\sum_{n} B_{n j-1} a_{n j-1} e^{-i q_{n j-1} (v t - L_{j-1})}. 
\end{align}

In order to satisfy (\ref{eq:z1}) for all time ($t$), we must have $A_{j} = 0$, and 
$\varphi_{j} = \varphi_{j-1} + \frac{2\pi}{m} \hat{N}_{j}$ for all $j$. Solving the latter set of linear equations, 
\begin{align} \label{eq:vp}
  \varphi_{j} = \frac{\pi}{m} \big( \hat{N}_{j} + \hat{N}_{j+1} - \hat{N}_{j-1} \big). 
\end{align}
As stated in the main text, here we assume that the edge modes are segments of a single boundary [hence we impose
Eq.~(\ref{eq:3md}) of the main text]. Therefore, we use a single wavevector $q_{n}$ and a single mode operator $a_{n}$ for all three chirals.
In this case, (\ref{eq:z2}) is satisfied for all $t$ only if 
\begin{align}
  B_{n j} a_{n} &= -B_{n j-1} e^{i q_{n} L_{j-1}} a_{n}, 
\end{align}
which in turn implies 
\begin{align}
  B_{n 3} a_{n} &= -B_{n 2} e^{i q_{n} L_{2}} a_{n} \nonumber \\
                &= B_{n 1} e^{i q_{n} (L_{2} + L_{1})} a_{n} \nonumber \\
                &= -B_{n 3} e^{i q_{n} \ell} a_{n} , \label{eq:QQ}
\end{align}
where $\ell = \sum_{j} L_{j}$ is the total length of the three chirals. Eq.~(\ref{eq:QQ}) naturally
leads to the quantization condition for the wavevectors, $q_{n} = (2n - 1) \frac{\pi}{\ell}$. Using the 
mode expansion in Eq.~(\ref{eq:2md}) of the main text (along with the standard bosonic commutations for $a_{n}$), we find that 
\begin{align}
  B_{n j}^{2} = \frac{2 \pi}{m \ell} \frac{1}{q_{n}}. \label{eq:bn}
\end{align}
Collecting everything together, we deduce that the chiral modes must satisfy,
\begin{align} \label{eq:pme1}
  \phi_1 (x,t) &= \varphi_{1} + \sqrt{\frac{2}{m}} \sum_{n > 0} \bigg[ \frac{a_{n}}{\sqrt{2n-1}}
  e^{-i q_{n} (vt - x)} + \text{h.c.} \bigg], \\
  \phi_2 (x,t) &= -\phi_1 (x + L_1,t) + \varphi_{1} + \varphi_{2}, \\
  \phi_3 (x,t) &= -\phi_2 (x + L_2,t) + \varphi_{2} + \varphi_{3}. \label{eq:pme3}
\end{align}
Similarly, using (\ref{eq:3md}) of the main text, we obtain the commutation relations between the zero mode operators,
\begin{align}
  [\varphi_{2}, \varphi_{1}] = [\varphi_{3}, \varphi_{2}] = \frac{2\pi i}{m} \, \text{ and } \,
  [\varphi_{1}, \varphi_{3}] = 0. \label{eq:zmcm1}
\end{align}
Using (\ref{eq:vp}) in (\ref{eq:zmcm1}), we find that the only non-zero commutation relation between the $\hat{N}$ operators is 
\begin{align} \label{eq:zmcm2}
  [\hat{N}_{3}, \hat{N}_{2}] = \frac{i m}{\pi}. 
\end{align}

Next, we use the mode expansion derived above in the Hamiltonian of the trijunction, $H = H_{\text{edge}} + H_{s} + H_{t}$ 
[the three components are defined in Eqs.~(\ref{eq:1md}), (\ref{eq:4md}) and (\ref{eq:5md}) of the main text]. 
Since $H_{\text{edge}}$ depends only on $\partial_{x} \phi_{j}$, we do not expect the zero mode operators ($\varphi_{j}$) 
to affect $H$. Indeed, using (\ref{eq:pme1}--\ref{eq:pme3}) we find,
\begin{align}
  H = \sum_{n > 0} v q_{n} \big(a_{n}^{\dagger} a_{n} + \frac{1}{2}\big).
\end{align}
In order to show the existence of a localized zero-energy excitation, we need to construct a physical operator 
(say, $\tilde{\Gamma}_{0}$) which commutes with $H$. Here, ``physical'' means that $\tilde{\Gamma}_{0}$ is only 
composed of operators corresponding to the local excitations of the effective theory. The low-energy dynamics of 
the junction is controlled by the bosonic excitations (described by $a_{n}$ and $a_{n}^{\dagger}$) and the anyonic 
quasihole/quasiparticle excitations (described by $e^{\mp i\phi_{j} (x)}$). Since the bosonic excitations have finite 
energy, any zero-energy excitation may only be composed of the vertex operators corresponding to quasihole/quasiparticle 
excitations. We begin the task of writing an explicit form of $\tilde{\Gamma}_{0}$ by noting that  
\begin{align}
  &[H, \phi_{j} (x)] = \sum_{n_{1}, n_{2} > 0} v q_{n_{2}} \sqrt{\frac{2}{m}} \frac{1}{\sqrt{2n_{1} - 1}} \times \nonumber \\
  &\,\,\,\,\,\,\,\,\,\,\,\,\,\,\,\,\,\,\,\,\,\,\,\,\, 
  \big[ a_{n_{2}}^{\dagger} a_{n_{2}}, a_{n_{1}} e^{-i q_{n_{1}} (vt - x)} + \text{h.c.} \big] \nonumber \\
  &= -v \sum_{n > 0} \sqrt{\frac{2}{m}} \frac{q_{n}}{\sqrt{2n - 1}} 
  \bigg[ a_{n} e^{-i q_{n} (vt - x)} - \text{h.c.} \bigg] \nonumber \\
  &= i v \partial_{x} \phi_{j} (x) . \label{eq:ff1}
\end{align}
Using (\ref{eq:ff1}) recursively, we find  
\begin{align}
  \big[H, \big(\phi_{j} (x) \big)^{n} \big] &=
  \big[H, \big(\phi_{j} (x) \big)^{n-1} \big] \phi_{j} (x) \nonumber \\ 
  &+ \big(\phi_{j} (x) \big)^{n-1} \big[H, \phi_{j} (x) \big] \nonumber \\
  &= i v \partial_{x} \big( \phi_{j} (x) \big)^{n}. \label{eq:ff2}
\end{align}
(\ref{eq:ff2}) may then be used to show that  
\begin{align}
  \big[H, e^{\phi_{j} (x) } \big] &= i v \partial_{x} \big( e^{\phi_{j} (x)} \big). \label{eq:ff3} 
\end{align}
Noting that the right hand side of (\ref{eq:ff3}) is a total derivative, we have, 
\begin{align} \label{eq:ff4}
  \bigg[H, \int_{0}^{L_{j}} dx \, e^{\phi_{j} (x)} \bigg] = i v \bigg( e^{i \phi_{j} (L_{j})} - e^{i \phi_{j} (0)} \bigg). 
\end{align}
Since $H$ commutes with $\varphi_{j}$, we rewrite (\ref{eq:ff4}) as, 
\begin{align} 
  \bigg[H, \int_{0}^{L_{j}} &dx \, e^{-i \varphi_{j}} e^{\phi_{j} (x)} \bigg] = 
  i v \bigg( e^{-i \varphi_{j}} e^{i \phi_{j} (L_{j})} - e^{-i \varphi_{j}} e^{i \phi_{j} (0)} \bigg) \nonumber \\
  &= i v \bigg( e^{i (\phi_{j} (L_{j}) - \varphi_{j})} - e^{i (\phi_{j} (0) - \varphi_{j})} \bigg) . \label{eq:ff5}
\end{align}
In the second equality above, we used the fact that $[\varphi_{j}, \phi_{j} (x)] = 0$. 
Employing (\ref{eq:bc}) and (\ref{eq:pme1}--\ref{eq:pme3}), it is trivial to show that,
\begin{align}
  \phi_{j} (0, t) - \varphi_{j} = - \big[ \phi_{j-1} (L_{j-1}, t) - \varphi_{j-1} \big] . \label{eq:ff6}
\end{align}
Finally, using (\ref{eq:ff5}) together with (\ref{eq:ff6}), we find   
\begin{align} \label{eq:ff7}
  \bigg[H, \sum_{j = 1}^{3} \int_{0}^{L_{j}} &dx \, \big( e^{-i \varphi_{j}} e^{\phi_{j} (x)} + \text{h.c.} \big) \bigg] = 0. 
\end{align}
We emphasize that Eq.~(\ref{eq:ff7}) is only valid in the case of an odd number of superconducting segments. 
In the case of an even-legged junction, $A_{j} \neq 0$, and the Hamiltonian would not be independent of $\hat{N}_{j}$. An equation 
of the form (\ref{eq:ff7}) would not be valid in such a situation. 

Even though the operator in (\ref{eq:ff7}) commutes with $H$, we have not reached our goal yet, since 
$e^{\mp i \varphi_{j}}$ (which is a crucial part of this operator) does not describe any physical excitation of 
the Hilbert space. In order to render it physical, we note that (\ref{eq:ff6}) may be written as   
\begin{align}
  \varphi_{j} + \varphi_{j-1} = \phi_{j} (0, t) + \phi_{j-1} (L_{j-1}, t) . \label{eq:ff8}
\end{align}
(\ref{eq:ff8}) motivates us to consider a family of operators of the form    
\begin{align}
  \tilde{\Gamma}_{0,k} = e^{i \varphi_{k}} \sum_{j = 1}^{3} \int_{0}^{L_{j}} 
  &dx \, \big( e^{-i \varphi_{j}} e^{\phi_{j} (x)} + \text{h.c.} \big), 
\end{align}
where $k = 1, 2, 3$. Clearly, they satisfy $[H, \tilde{\Gamma}_{0,k}] = 0$. Additionally, up to trivial phase factors, 
the zero-mode operators enter $\tilde{\Gamma}_{0,k}$ in one of three forms: $e^{i (\varphi_{j} + \varphi_{k})}$,   
$e^{2 i \varphi_{k}}$, or $e^{i (\varphi_{k} - \varphi_{j})}$. According to (\ref{eq:ff8}), $e^{i (\varphi_{j} + \varphi_{k})}$ 
represents a local pair of quasihole and quasiparticle excitations at the boundary of one of the
superconducting segments. The second and third forms may also be interpreted in a similar way, since (up to trivial phase factors) 
$e^{2 i \varphi_{k}} = e^{i (\varphi_{k} + \varphi_{k+1})} e^{i (\varphi_{k} + \varphi_{k-1})} 
e^{-i (\varphi_{k+1} + \varphi_{k-1})}$, and $e^{i (\varphi_{k} - \varphi_{j})} = e^{i (\varphi_{k} + \varphi_{k+1})} 
e^{-i (\varphi_{j} + \varphi_{k+1})}$. Hence, we conclude that $\tilde{\Gamma}_{0, k}$ indeed represent physical
zero-energy excitations of the system. Note, however, that $\tilde{\Gamma}_{0,k}$ for different $k$ are not independent of each other. 
Indeed, they are related to each other by   
\begin{align}
  \tilde{\Gamma}_{0,k_{2}} = e^{i (\varphi_{k_{2}} - \varphi_{k_{1}})} \tilde{\Gamma}_{0,k_{1}}, 
\end{align}
and as we mentioned above, $e^{i (\varphi_{k_{2}} - \varphi_{k_{1}})}$ may be decomposed in terms of physical operators. 
In other words, the three operators represent the same zero-energy excitation, but with different dressing of quasihole-quasiparticle pairs. 
Therefore, in Eq.~(\ref{eq:19md}) of the main text, we define a single zero-energy excitation ($\tilde{\Gamma}_{0}$), which without loss of generality 
may be identified with ($\tilde{\Gamma}_{0,1}$).


\begin{thebibliography}{73}%
\makeatletter
\providecommand \@ifxundefined [1]{%
 \@ifx{#1\undefined}
}%
\providecommand \@ifnum [1]{%
 \ifnum #1\expandafter \@firstoftwo
 \else \expandafter \@secondoftwo
 \fi
}%
\providecommand \@ifx [1]{%
 \ifx #1\expandafter \@firstoftwo
 \else \expandafter \@secondoftwo
 \fi
}%
\providecommand \natexlab [1]{#1}%
\providecommand \enquote  [1]{``#1''}%
\providecommand \bibnamefont  [1]{#1}%
\providecommand \bibfnamefont [1]{#1}%
\providecommand \citenamefont [1]{#1}%
\providecommand \href@noop [0]{\@secondoftwo}%
\providecommand \href [0]{\begingroup \@sanitize@url \@href}%
\providecommand \@href[1]{\@@startlink{#1}\@@href}%
\providecommand \@@href[1]{\endgroup#1\@@endlink}%
\providecommand \@sanitize@url [0]{\catcode `\\12\catcode `\$12\catcode
  `\&12\catcode `\#12\catcode `\^12\catcode `\_12\catcode `\%12\relax}%
\providecommand \@@startlink[1]{}%
\providecommand \@@endlink[0]{}%
\providecommand \url  [0]{\begingroup\@sanitize@url \@url }%
\providecommand \@url [1]{\endgroup\@href {#1}{\urlprefix }}%
\providecommand \urlprefix  [0]{URL }%
\providecommand \Eprint [0]{\href }%
\providecommand \doibase [0]{https://doi.org/}%
\providecommand \selectlanguage [0]{\@gobble}%
\providecommand \bibinfo  [0]{\@secondoftwo}%
\providecommand \bibfield  [0]{\@secondoftwo}%
\providecommand \translation [1]{[#1]}%
\providecommand \BibitemOpen [0]{}%
\providecommand \bibitemStop [0]{}%
\providecommand \bibitemNoStop [0]{.\EOS\space}%
\providecommand \EOS [0]{\spacefactor3000\relax}%
\providecommand \BibitemShut  [1]{\csname bibitem#1\endcsname}%
\let\auto@bib@innerbib\@empty
\bibitem [{\citenamefont {Alicea}\ and\ \citenamefont
  {Fendley}(2016)}]{AliceaReview}%
  \BibitemOpen
  \bibfield  {author} {\bibinfo {author} {\bibfnamefont {J.}~\bibnamefont
  {Alicea}}\ and\ \bibinfo {author} {\bibfnamefont {P.}~\bibnamefont
  {Fendley}},\ }\bibfield  {title} {\bibinfo {title} {Topological phases with
  parafermions: Theory and blueprints},\ }\href
  {https://doi.org/10.1146/annurev-conmatphys-031115-011336} {\bibfield
  {journal} {\bibinfo  {journal} {Annu. Rev. Condens. Matter Phys.}\ }\textbf
  {\bibinfo {volume} {7}},\ \bibinfo {pages} {119} (\bibinfo {year}
  {2016})}\BibitemShut {NoStop}%
\bibitem [{\citenamefont {Clarke}\ \emph
  {et~al.}(2013{\natexlab{a}})\citenamefont {Clarke}, \citenamefont {Alicea},\
  and\ \citenamefont {Shtengel}}]{Alicea1}%
  \BibitemOpen
  \bibfield  {author} {\bibinfo {author} {\bibfnamefont {D.~J.}\ \bibnamefont
  {Clarke}}, \bibinfo {author} {\bibfnamefont {J.}~\bibnamefont {Alicea}},\
  and\ \bibinfo {author} {\bibfnamefont {K.}~\bibnamefont {Shtengel}},\ }\href
  {https://doi.org/10.1038/ncomms2340} {\bibfield  {journal} {\bibinfo
  {journal} {Nat. Commun.}\ }\textbf {\bibinfo {volume} {4}},\ \bibinfo {pages}
  {1348} (\bibinfo {year} {2013}{\natexlab{a}})}\BibitemShut {NoStop}%
\bibitem [{\citenamefont {Lindner}\ \emph {et~al.}(2012)\citenamefont
  {Lindner}, \citenamefont {Berg}, \citenamefont {Refael},\ and\ \citenamefont
  {Stern}}]{Stern1}%
  \BibitemOpen
  \bibfield  {author} {\bibinfo {author} {\bibfnamefont {N.~H.}\ \bibnamefont
  {Lindner}}, \bibinfo {author} {\bibfnamefont {E.}~\bibnamefont {Berg}},
  \bibinfo {author} {\bibfnamefont {G.}~\bibnamefont {Refael}},\ and\ \bibinfo
  {author} {\bibfnamefont {A.}~\bibnamefont {Stern}},\ }\bibfield  {title}
  {\bibinfo {title} {Fractionalizing {M}ajorana fermions: Non-{A}belian
  statistics on the edges of {A}belian quantum {H}all states},\ }\href
  {https://doi.org/10.1103/PhysRevX.2.041002} {\bibfield  {journal} {\bibinfo
  {journal} {Phys. Rev. X}\ }\textbf {\bibinfo {volume} {2}},\ \bibinfo {pages}
  {041002} (\bibinfo {year} {2012})}\BibitemShut {NoStop}%
\bibitem [{\citenamefont {Santos}\ and\ \citenamefont
  {Hughes}(2017)}]{Luiz2017}%
  \BibitemOpen
  \bibfield  {author} {\bibinfo {author} {\bibfnamefont {L.~H.}\ \bibnamefont
  {Santos}}\ and\ \bibinfo {author} {\bibfnamefont {T.~L.}\ \bibnamefont
  {Hughes}},\ }\bibfield  {title} {\bibinfo {title} {Parafermionic wires at the
  interface of chiral topological states},\ }\href
  {https://doi.org/10.1103/PhysRevLett.118.136801} {\bibfield  {journal}
  {\bibinfo  {journal} {Phys. Rev. Lett.}\ }\textbf {\bibinfo {volume} {118}},\
  \bibinfo {pages} {136801} (\bibinfo {year} {2017})}\BibitemShut {NoStop}%
\bibitem [{\citenamefont {Mong}\ \emph {et~al.}(2014)\citenamefont {Mong},
  \citenamefont {Clarke}, \citenamefont {Alicea}, \citenamefont {Lindner},
  \citenamefont {Fendley}, \citenamefont {Nayak}, \citenamefont {Oreg},
  \citenamefont {Stern}, \citenamefont {Berg}, \citenamefont {Shtengel},\ and\
  \citenamefont {Fisher}}]{Stern2}%
  \BibitemOpen
  \bibfield  {author} {\bibinfo {author} {\bibfnamefont {R.~S.~K.}\
  \bibnamefont {Mong}}, \bibinfo {author} {\bibfnamefont {D.~J.}\ \bibnamefont
  {Clarke}}, \bibinfo {author} {\bibfnamefont {J.}~\bibnamefont {Alicea}},
  \bibinfo {author} {\bibfnamefont {N.~H.}\ \bibnamefont {Lindner}}, \bibinfo
  {author} {\bibfnamefont {P.}~\bibnamefont {Fendley}}, \bibinfo {author}
  {\bibfnamefont {C.}~\bibnamefont {Nayak}}, \bibinfo {author} {\bibfnamefont
  {Y.}~\bibnamefont {Oreg}}, \bibinfo {author} {\bibfnamefont {A.}~\bibnamefont
  {Stern}}, \bibinfo {author} {\bibfnamefont {E.}~\bibnamefont {Berg}},
  \bibinfo {author} {\bibfnamefont {K.}~\bibnamefont {Shtengel}},\ and\
  \bibinfo {author} {\bibfnamefont {M.~P.~A.}\ \bibnamefont {Fisher}},\
  }\bibfield  {title} {\bibinfo {title} {Universal topological quantum
  computation from a superconductor-{A}belian quantum {H}all heterostructure},\
  }\href {https://doi.org/10.1103/PhysRevX.4.011036} {\bibfield  {journal}
  {\bibinfo  {journal} {Phys. Rev. X}\ }\textbf {\bibinfo {volume} {4}},\
  \bibinfo {pages} {011036} (\bibinfo {year} {2014})}\BibitemShut {NoStop}%
\bibitem [{\citenamefont {Alicea}\ and\ \citenamefont
  {Stern}(2015)}]{Fibonacci2}%
  \BibitemOpen
  \bibfield  {author} {\bibinfo {author} {\bibfnamefont {J.}~\bibnamefont
  {Alicea}}\ and\ \bibinfo {author} {\bibfnamefont {A.}~\bibnamefont {Stern}},\
  }\bibfield  {title} {\bibinfo {title} {Designer non-{A}belian anyon
  platforms: from {M}ajorana to {F}ibonacci},\ }\href
  {https://doi.org/10.1088/0031-8949/2015/T164/014006} {\bibfield  {journal}
  {\bibinfo  {journal} {Phys. Scr.}\ }\textbf {\bibinfo {volume} {2015}},\
  \bibinfo {pages} {014006} (\bibinfo {year} {2015})}\BibitemShut {NoStop}%
\bibitem [{\citenamefont {Clarke}\ \emph
  {et~al.}(2013{\natexlab{b}})\citenamefont {Clarke}, \citenamefont {Alicea},\
  and\ \citenamefont {Shtengel}}]{Alicea2}%
  \BibitemOpen
  \bibfield  {author} {\bibinfo {author} {\bibfnamefont {D.~J.}\ \bibnamefont
  {Clarke}}, \bibinfo {author} {\bibfnamefont {J.}~\bibnamefont {Alicea}},\
  and\ \bibinfo {author} {\bibfnamefont {K.}~\bibnamefont {Shtengel}},\ }\href
  {https://doi.org/10.1038/nphys3114} {\bibfield  {journal} {\bibinfo
  {journal} {Nat. Phys.}\ }\textbf {\bibinfo {volume} {10}},\ \bibinfo {pages}
  {877} (\bibinfo {year} {2014}{\natexlab{b}})}\BibitemShut {NoStop}%
\bibitem [{\citenamefont {Cheng}(2012)}]{App1}%
  \BibitemOpen
  \bibfield  {author} {\bibinfo {author} {\bibfnamefont {M.}~\bibnamefont
  {Cheng}},\ }\bibfield  {title} {\bibinfo {title} {Superconducting proximity
  effect on the edge of fractional topological insulators},\ }\href
  {https://doi.org/10.1103/PhysRevB.86.195126} {\bibfield  {journal} {\bibinfo
  {journal} {Phys. Rev. B}\ }\textbf {\bibinfo {volume} {86}},\ \bibinfo
  {pages} {195126} (\bibinfo {year} {2012})}\BibitemShut {NoStop}%
\bibitem [{\citenamefont {Burrello}\ \emph {et~al.}(2013)\citenamefont
  {Burrello}, \citenamefont {van Heck},\ and\ \citenamefont {Cobanera}}]{App2}%
  \BibitemOpen
  \bibfield  {author} {\bibinfo {author} {\bibfnamefont {M.}~\bibnamefont
  {Burrello}}, \bibinfo {author} {\bibfnamefont {B.}~\bibnamefont {van Heck}},\
  and\ \bibinfo {author} {\bibfnamefont {E.}~\bibnamefont {Cobanera}},\
  }\bibfield  {title} {\bibinfo {title} {Topological phases in two-dimensional
  arrays of parafermionic zero modes},\ }\href
  {https://doi.org/10.1103/PhysRevB.87.195422} {\bibfield  {journal} {\bibinfo
  {journal} {Phys. Rev. B}\ }\textbf {\bibinfo {volume} {87}},\ \bibinfo
  {pages} {195422} (\bibinfo {year} {2013})}\BibitemShut {NoStop}%
\bibitem [{\citenamefont {Vaezi}(2013)}]{App3}%
  \BibitemOpen
  \bibfield  {author} {\bibinfo {author} {\bibfnamefont {A.}~\bibnamefont
  {Vaezi}},\ }\bibfield  {title} {\bibinfo {title} {Fractional topological
  superconductor with fractionalized {M}ajorana fermions},\ }\href
  {https://doi.org/10.1103/PhysRevB.87.035132} {\bibfield  {journal} {\bibinfo
  {journal} {Phys. Rev. B}\ }\textbf {\bibinfo {volume} {87}},\ \bibinfo
  {pages} {035132} (\bibinfo {year} {2013})}\BibitemShut {NoStop}%
\bibitem [{\citenamefont {Zhang}\ and\ \citenamefont {Kane}(2014)}]{Zhang1}%
  \BibitemOpen
  \bibfield  {author} {\bibinfo {author} {\bibfnamefont {F.}~\bibnamefont
  {Zhang}}\ and\ \bibinfo {author} {\bibfnamefont {C.~L.}\ \bibnamefont
  {Kane}},\ }\bibfield  {title} {\bibinfo {title} {Time-reversal-invariant
  $\mathbb{Z}_{4}$ fractional {J}osephson effect},\ }\href
  {https://doi.org/10.1103/PhysRevLett.113.036401} {\bibfield  {journal}
  {\bibinfo  {journal} {Phys. Rev. Lett.}\ }\textbf {\bibinfo {volume} {113}},\
  \bibinfo {pages} {036401} (\bibinfo {year} {2014})}\BibitemShut {NoStop}%
\bibitem [{\citenamefont {Cheng}\ and\ \citenamefont {Lutchyn}(2015)}]{App4}%
  \BibitemOpen
  \bibfield  {author} {\bibinfo {author} {\bibfnamefont {M.}~\bibnamefont
  {Cheng}}\ and\ \bibinfo {author} {\bibfnamefont {R.}~\bibnamefont
  {Lutchyn}},\ }\bibfield  {title} {\bibinfo {title} {Fractional {J}osephson
  effect in number-conserving systems},\ }\href
  {https://doi.org/10.1103/PhysRevB.92.134516} {\bibfield  {journal} {\bibinfo
  {journal} {Phys. Rev. B}\ }\textbf {\bibinfo {volume} {92}},\ \bibinfo
  {pages} {134516} (\bibinfo {year} {2015})}\BibitemShut {NoStop}%
\bibitem [{\citenamefont {Kim}\ \emph {et~al.}(2017)\citenamefont {Kim},
  \citenamefont {Clarke},\ and\ \citenamefont {Lutchyn}}]{App6}%
  \BibitemOpen
  \bibfield  {author} {\bibinfo {author} {\bibfnamefont {Y.}~\bibnamefont
  {Kim}}, \bibinfo {author} {\bibfnamefont {D.~J.}\ \bibnamefont {Clarke}},\
  and\ \bibinfo {author} {\bibfnamefont {R.~M.}\ \bibnamefont {Lutchyn}},\
  }\bibfield  {title} {\bibinfo {title} {Coulomb blockade in fractional
  topological superconductors},\ }\href
  {https://doi.org/10.1103/PhysRevB.96.041123} {\bibfield  {journal} {\bibinfo
  {journal} {Phys. Rev. B}\ }\textbf {\bibinfo {volume} {96}},\ \bibinfo
  {pages} {041123} (\bibinfo {year} {2017})}\BibitemShut {NoStop}%
\bibitem [{\citenamefont {Beri}\ and\ \citenamefont {Cooper}(2012)}]{App7}%
  \BibitemOpen
  \bibfield  {author} {\bibinfo {author} {\bibfnamefont {B.}~\bibnamefont
  {Beri}}\ and\ \bibinfo {author} {\bibfnamefont {N.~R.}\ \bibnamefont
  {Cooper}},\ }\bibfield  {title} {\bibinfo {title} {Topological {K}ondo effect
  with {M}ajorana fermions},\ }\href
  {https://doi.org/10.1103/PhysRevLett.109.156803} {\bibfield  {journal}
  {\bibinfo  {journal} {Phys. Rev. Lett.}\ }\textbf {\bibinfo {volume} {109}},\
  \bibinfo {pages} {156803} (\bibinfo {year} {2012})}\BibitemShut {NoStop}%
\bibitem [{\citenamefont {Altland}\ and\ \citenamefont {Egger}(2013)}]{App8}%
  \BibitemOpen
  \bibfield  {author} {\bibinfo {author} {\bibfnamefont {A.}~\bibnamefont
  {Altland}}\ and\ \bibinfo {author} {\bibfnamefont {R.}~\bibnamefont
  {Egger}},\ }\bibfield  {title} {\bibinfo {title} {Multiterminal
  {C}oulomb-{M}ajorana junction},\ }\href
  {https://doi.org/10.1103/PhysRevLett.110.196401} {\bibfield  {journal}
  {\bibinfo  {journal} {Phys. Rev. Lett.}\ }\textbf {\bibinfo {volume} {110}},\
  \bibinfo {pages} {196401} (\bibinfo {year} {2013})}\BibitemShut {NoStop}%
\bibitem [{\citenamefont {Beri}(2013)}]{App9}%
  \BibitemOpen
  \bibfield  {author} {\bibinfo {author} {\bibfnamefont {B.}~\bibnamefont
  {Beri}},\ }\bibfield  {title} {\bibinfo {title} {{M}ajorana-{K}lein
  hybridization in topological superconductor junctions},\ }\href
  {https://doi.org/10.1103/PhysRevLett.110.216803} {\bibfield  {journal}
  {\bibinfo  {journal} {Phys. Rev. Lett.}\ }\textbf {\bibinfo {volume} {110}},\
  \bibinfo {pages} {216803} (\bibinfo {year} {2013})}\BibitemShut {NoStop}%
\bibitem [{\citenamefont {Snizhko}\ \emph
  {et~al.}(2018{\natexlab{a}})\citenamefont {Snizhko}, \citenamefont
  {Buccheri}, \citenamefont {Egger},\ and\ \citenamefont {Gefen}}]{App10}%
  \BibitemOpen
  \bibfield  {author} {\bibinfo {author} {\bibfnamefont {K.}~\bibnamefont
  {Snizhko}}, \bibinfo {author} {\bibfnamefont {F.}~\bibnamefont {Buccheri}},
  \bibinfo {author} {\bibfnamefont {R.}~\bibnamefont {Egger}},\ and\ \bibinfo
  {author} {\bibfnamefont {Y.}~\bibnamefont {Gefen}},\ }\bibfield  {title}
  {\bibinfo {title} {Parafermionic generalization of the topological {K}ondo
  effect},\ }\href {https://doi.org/10.1103/PhysRevB.97.235139} {\bibfield
  {journal} {\bibinfo  {journal} {Phys. Rev. B}\ }\textbf {\bibinfo {volume}
  {97}},\ \bibinfo {pages} {235139} (\bibinfo {year}
  {2018}{\natexlab{a}})}\BibitemShut {NoStop}%
\bibitem [{\citenamefont {Gau}\ \emph {et~al.}(2018)\citenamefont {Gau},
  \citenamefont {Plugge},\ and\ \citenamefont {Egger}}]{App11}%
  \BibitemOpen
  \bibfield  {author} {\bibinfo {author} {\bibfnamefont {M.}~\bibnamefont
  {Gau}}, \bibinfo {author} {\bibfnamefont {S.}~\bibnamefont {Plugge}},\ and\
  \bibinfo {author} {\bibfnamefont {R.}~\bibnamefont {Egger}},\ }\bibfield
  {title} {\bibinfo {title} {Quantum transport in coupled {M}ajorana box
  systems},\ }\href {https://doi.org/10.1103/PhysRevB.97.184506} {\bibfield
  {journal} {\bibinfo  {journal} {Phys. Rev. B}\ }\textbf {\bibinfo {volume}
  {97}},\ \bibinfo {pages} {184506} (\bibinfo {year} {2018})}\BibitemShut
  {NoStop}%
\bibitem [{\citenamefont {Herasymenko}\ \emph
  {et~al.}(2018{\natexlab{a}})\citenamefont {Herasymenko}, \citenamefont
  {Snizhko},\ and\ \citenamefont {Gefen}}]{App12}%
  \BibitemOpen
  \bibfield  {author} {\bibinfo {author} {\bibfnamefont {Y.}~\bibnamefont
  {Herasymenko}}, \bibinfo {author} {\bibfnamefont {K.}~\bibnamefont
  {Snizhko}},\ and\ \bibinfo {author} {\bibfnamefont {Y.}~\bibnamefont
  {Gefen}},\ }\bibfield  {title} {\bibinfo {title} {Universal quantum noise in
  adiabatic pumping},\ }\href {https://doi.org/10.1103/PhysRevLett.120.226802}
  {\bibfield  {journal} {\bibinfo  {journal} {Phys. Rev. Lett.}\ }\textbf
  {\bibinfo {volume} {120}},\ \bibinfo {pages} {226802} (\bibinfo {year}
  {2018}{\natexlab{a}})}\BibitemShut {NoStop}%
\bibitem [{\citenamefont {Pedder}\ \emph {et~al.}(2017)\citenamefont {Pedder},
  \citenamefont {Meng}, \citenamefont {Tiwari},\ and\ \citenamefont
  {Schmidt}}]{Tiwari17}%
  \BibitemOpen
  \bibfield  {author} {\bibinfo {author} {\bibfnamefont {C.~J.}\ \bibnamefont
  {Pedder}}, \bibinfo {author} {\bibfnamefont {T.}~\bibnamefont {Meng}},
  \bibinfo {author} {\bibfnamefont {R.~P.}\ \bibnamefont {Tiwari}},\ and\
  \bibinfo {author} {\bibfnamefont {T.~L.}\ \bibnamefont {Schmidt}},\
  }\bibfield  {title} {\bibinfo {title} {Missing shapiro steps and the
  $8\ensuremath{\pi}$-periodic {J}osephson effect in interacting helical
  electron systems},\ }\href {https://doi.org/10.1103/PhysRevB.96.165429}
  {\bibfield  {journal} {\bibinfo  {journal} {Phys. Rev. B}\ }\textbf {\bibinfo
  {volume} {96}},\ \bibinfo {pages} {165429} (\bibinfo {year}
  {2017})}\BibitemShut {NoStop}%
\bibitem [{\citenamefont {Svetogorov}\ \emph {et~al.}(2021)\citenamefont
  {Svetogorov}, \citenamefont {Loss},\ and\ \citenamefont
  {Klinovaja}}]{Jelena2021}%
  \BibitemOpen
  \bibfield  {author} {\bibinfo {author} {\bibfnamefont {A.~E.}\ \bibnamefont
  {Svetogorov}}, \bibinfo {author} {\bibfnamefont {D.}~\bibnamefont {Loss}},\
  and\ \bibinfo {author} {\bibfnamefont {J.}~\bibnamefont {Klinovaja}},\
  }\bibfield  {title} {\bibinfo {title} {Insulating regime of an underdamped
  current-biased {J}osephson junction supporting $\mathbb{Z}_{3}$ and
  $\mathbb{Z}_{4}$ parafermions},\ }\href
  {https://doi.org/10.1103/PhysRevB.103.L180505} {\bibfield  {journal}
  {\bibinfo  {journal} {Phys. Rev. B}\ }\textbf {\bibinfo {volume} {103}},\
  \bibinfo {pages} {L180505} (\bibinfo {year} {2021})}\BibitemShut {NoStop}%
\bibitem [{\citenamefont {Nayak}\ \emph {et~al.}(2008)\citenamefont {Nayak},
  \citenamefont {Simon}, \citenamefont {Stern}, \citenamefont {Freedman},\ and\
  \citenamefont {Das~Sarma}}]{NayakRMP}%
  \BibitemOpen
  \bibfield  {author} {\bibinfo {author} {\bibfnamefont {C.}~\bibnamefont
  {Nayak}}, \bibinfo {author} {\bibfnamefont {S.~H.}\ \bibnamefont {Simon}},
  \bibinfo {author} {\bibfnamefont {A.}~\bibnamefont {Stern}}, \bibinfo
  {author} {\bibfnamefont {M.}~\bibnamefont {Freedman}},\ and\ \bibinfo
  {author} {\bibfnamefont {S.}~\bibnamefont {Das~Sarma}},\ }\bibfield  {title}
  {\bibinfo {title} {Non-{A}belian anyons and topological quantum
  computation},\ }\href {https://doi.org/10.1103/RevModPhys.80.1083} {\bibfield
   {journal} {\bibinfo  {journal} {Rev. Mod. Phys.}\ }\textbf {\bibinfo
  {volume} {80}},\ \bibinfo {pages} {1083} (\bibinfo {year}
  {2008})}\BibitemShut {NoStop}%
\bibitem [{\citenamefont {Bonderson}\ \emph {et~al.}(2008)\citenamefont
  {Bonderson}, \citenamefont {Freedman},\ and\ \citenamefont {Nayak}}]{QIC1}%
  \BibitemOpen
  \bibfield  {author} {\bibinfo {author} {\bibfnamefont {P.}~\bibnamefont
  {Bonderson}}, \bibinfo {author} {\bibfnamefont {M.}~\bibnamefont
  {Freedman}},\ and\ \bibinfo {author} {\bibfnamefont {C.}~\bibnamefont
  {Nayak}},\ }\bibfield  {title} {\bibinfo {title} {Measurement-only
  topological quantum computation},\ }\href
  {https://doi.org/10.1103/PhysRevLett.101.010501} {\bibfield  {journal}
  {\bibinfo  {journal} {Phys. Rev. Lett.}\ }\textbf {\bibinfo {volume} {101}},\
  \bibinfo {pages} {010501} (\bibinfo {year} {2008})}\BibitemShut {NoStop}%
\bibitem [{\citenamefont {Zheng}\ \emph {et~al.}(2016)\citenamefont {Zheng},
  \citenamefont {Dua},\ and\ \citenamefont {Jiang}}]{QIC2}%
  \BibitemOpen
  \bibfield  {author} {\bibinfo {author} {\bibfnamefont {H.}~\bibnamefont
  {Zheng}}, \bibinfo {author} {\bibfnamefont {A.~D.}\ \bibnamefont {Dua}},\
  and\ \bibinfo {author} {\bibfnamefont {L.}~\bibnamefont {Jiang}},\ }\bibfield
   {title} {\bibinfo {title} {Measurement-only topological quantum computation
  without forced measurements},\ }\href
  {https://doi.org/10.1088/1367-2630/aa50bb} {\bibfield  {journal} {\bibinfo
  {journal} {New J. Phys.}\ }\textbf {\bibinfo {volume} {18}},\ \bibinfo
  {pages} {123087} (\bibinfo {year} {2016})}\BibitemShut {NoStop}%
\bibitem [{\citenamefont {Hutter}\ and\ \citenamefont {Loss}(2016)}]{QIC3}%
  \BibitemOpen
  \bibfield  {author} {\bibinfo {author} {\bibfnamefont {A.}~\bibnamefont
  {Hutter}}\ and\ \bibinfo {author} {\bibfnamefont {D.}~\bibnamefont {Loss}},\
  }\bibfield  {title} {\bibinfo {title} {Quantum computing with parafermions},\
  }\href {https://doi.org/10.1103/PhysRevB.93.125105} {\bibfield  {journal}
  {\bibinfo  {journal} {Phys. Rev. B}\ }\textbf {\bibinfo {volume} {93}},\
  \bibinfo {pages} {125105} (\bibinfo {year} {2016})}\BibitemShut {NoStop}%
\bibitem [{\citenamefont {Dua}\ \emph {et~al.}(2019)\citenamefont {Dua},
  \citenamefont {Malomed}, \citenamefont {Cheng},\ and\ \citenamefont
  {Jiang}}]{QIC4}%
  \BibitemOpen
  \bibfield  {author} {\bibinfo {author} {\bibfnamefont {A.}~\bibnamefont
  {Dua}}, \bibinfo {author} {\bibfnamefont {B.}~\bibnamefont {Malomed}},
  \bibinfo {author} {\bibfnamefont {M.}~\bibnamefont {Cheng}},\ and\ \bibinfo
  {author} {\bibfnamefont {L.}~\bibnamefont {Jiang}},\ }\bibfield  {title}
  {\bibinfo {title} {Universal quantum computing with parafermions assisted by
  a half-fluxon},\ }\href {https://doi.org/10.1103/PhysRevB.100.144508}
  {\bibfield  {journal} {\bibinfo  {journal} {Phys. Rev. B}\ }\textbf {\bibinfo
  {volume} {100}},\ \bibinfo {pages} {144508} (\bibinfo {year}
  {2019})}\BibitemShut {NoStop}%
\bibitem [{\citenamefont {Carmi}\ \emph {et~al.}(2019)\citenamefont {Carmi},
  \citenamefont {Herasymenko}, \citenamefont {Cohen},\ and\ \citenamefont
  {Snizhko}}]{QIC5}%
  \BibitemOpen
  \bibfield  {author} {\bibinfo {author} {\bibfnamefont {A.}~\bibnamefont
  {Carmi}}, \bibinfo {author} {\bibfnamefont {Y.}~\bibnamefont {Herasymenko}},
  \bibinfo {author} {\bibfnamefont {E.}~\bibnamefont {Cohen}},\ and\ \bibinfo
  {author} {\bibfnamefont {K.}~\bibnamefont {Snizhko}},\ }\bibfield  {title}
  {\bibinfo {title} {Bounds on nonlocal correlations in the presence of
  signaling and their application to topological zero modes},\ }\href
  {https://doi.org/10.1088/1367-2630/ab2f5b} {\bibfield  {journal} {\bibinfo
  {journal} {New J. Phys.}\ }\textbf {\bibinfo {volume} {21}},\ \bibinfo
  {pages} {073032} (\bibinfo {year} {2019})}\BibitemShut {NoStop}%
\bibitem [{\citenamefont {Groenendijk}\ \emph {et~al.}(2019)\citenamefont
  {Groenendijk}, \citenamefont {Calzona}, \citenamefont {Tschirhart},
  \citenamefont {Idrisov},\ and\ \citenamefont {Schmidt}}]{QIC6}%
  \BibitemOpen
  \bibfield  {author} {\bibinfo {author} {\bibfnamefont {S.}~\bibnamefont
  {Groenendijk}}, \bibinfo {author} {\bibfnamefont {A.}~\bibnamefont
  {Calzona}}, \bibinfo {author} {\bibfnamefont {H.}~\bibnamefont {Tschirhart}},
  \bibinfo {author} {\bibfnamefont {E.~G.}\ \bibnamefont {Idrisov}},\ and\
  \bibinfo {author} {\bibfnamefont {T.~L.}\ \bibnamefont {Schmidt}},\
  }\bibfield  {title} {\bibinfo {title} {Parafermion braiding in fractional
  quantum {H}all edge states with a finite chemical potential},\ }\href
  {https://doi.org/10.1103/PhysRevB.100.205424} {\bibfield  {journal} {\bibinfo
   {journal} {Phys. Rev. B}\ }\textbf {\bibinfo {volume} {100}},\ \bibinfo
  {pages} {205424} (\bibinfo {year} {2019})}\BibitemShut {NoStop}%
\bibitem [{\citenamefont {Landau}\ \emph {et~al.}(2016)\citenamefont {Landau},
  \citenamefont {Plugge}, \citenamefont {Sela}, \citenamefont {Altland},
  \citenamefont {Albrecht},\ and\ \citenamefont {Egger}}]{QIC7}%
  \BibitemOpen
  \bibfield  {author} {\bibinfo {author} {\bibfnamefont {L.~A.}\ \bibnamefont
  {Landau}}, \bibinfo {author} {\bibfnamefont {S.}~\bibnamefont {Plugge}},
  \bibinfo {author} {\bibfnamefont {E.}~\bibnamefont {Sela}}, \bibinfo {author}
  {\bibfnamefont {A.}~\bibnamefont {Altland}}, \bibinfo {author} {\bibfnamefont
  {S.~M.}\ \bibnamefont {Albrecht}},\ and\ \bibinfo {author} {\bibfnamefont
  {R.}~\bibnamefont {Egger}},\ }\bibfield  {title} {\bibinfo {title} {Towards
  realistic implementations of a {M}ajorana surface code},\ }\href
  {https://doi.org/10.1103/PhysRevLett.116.050501} {\bibfield  {journal}
  {\bibinfo  {journal} {Phys. Rev. Lett.}\ }\textbf {\bibinfo {volume} {116}},\
  \bibinfo {pages} {050501} (\bibinfo {year} {2016})}\BibitemShut {NoStop}%
\bibitem [{\citenamefont {Knapp}\ \emph {et~al.}(2020)\citenamefont {Knapp},
  \citenamefont {V\"ayrynen},\ and\ \citenamefont {Lutchyn}}]{QIC8}%
  \BibitemOpen
  \bibfield  {author} {\bibinfo {author} {\bibfnamefont {C.}~\bibnamefont
  {Knapp}}, \bibinfo {author} {\bibfnamefont {J.~I.}\ \bibnamefont
  {V\"ayrynen}},\ and\ \bibinfo {author} {\bibfnamefont {R.~M.}\ \bibnamefont
  {Lutchyn}},\ }\bibfield  {title} {\bibinfo {title} {Number-conserving
  analysis of measurement-based braiding with {M}ajorana zero modes},\ }\href
  {https://doi.org/10.1103/PhysRevB.101.125108} {\bibfield  {journal} {\bibinfo
   {journal} {Phys. Rev. B}\ }\textbf {\bibinfo {volume} {101}},\ \bibinfo
  {pages} {125108} (\bibinfo {year} {2020})}\BibitemShut {NoStop}%
\bibitem [{\citenamefont {Fidkowski}\ and\ \citenamefont
  {Kitaev}(2010)}]{Class1}%
  \BibitemOpen
  \bibfield  {author} {\bibinfo {author} {\bibfnamefont {L.}~\bibnamefont
  {Fidkowski}}\ and\ \bibinfo {author} {\bibfnamefont {A.}~\bibnamefont
  {Kitaev}},\ }\bibfield  {title} {\bibinfo {title} {Effects of interactions on
  the topological classification of free fermion systems},\ }\href
  {https://doi.org/10.1103/PhysRevB.81.134509} {\bibfield  {journal} {\bibinfo
  {journal} {Phys. Rev. B}\ }\textbf {\bibinfo {volume} {81}},\ \bibinfo
  {pages} {134509} (\bibinfo {year} {2010})}\BibitemShut {NoStop}%
\bibitem [{\citenamefont {Turner}\ \emph {et~al.}(2011)\citenamefont {Turner},
  \citenamefont {Pollmann},\ and\ \citenamefont {Berg}}]{Class2}%
  \BibitemOpen
  \bibfield  {author} {\bibinfo {author} {\bibfnamefont {A.~M.}\ \bibnamefont
  {Turner}}, \bibinfo {author} {\bibfnamefont {F.}~\bibnamefont {Pollmann}},\
  and\ \bibinfo {author} {\bibfnamefont {E.}~\bibnamefont {Berg}},\ }\bibfield
  {title} {\bibinfo {title} {Topological phases of one-dimensional fermions: An
  entanglement point of view},\ }\href
  {https://doi.org/10.1103/PhysRevB.83.075102} {\bibfield  {journal} {\bibinfo
  {journal} {Phys. Rev. B}\ }\textbf {\bibinfo {volume} {83}},\ \bibinfo
  {pages} {075102} (\bibinfo {year} {2011})}\BibitemShut {NoStop}%
\bibitem [{\citenamefont {Chen}\ \emph {et~al.}(2011)\citenamefont {Chen},
  \citenamefont {Gu},\ and\ \citenamefont {Wen}}]{Class3}%
  \BibitemOpen
  \bibfield  {author} {\bibinfo {author} {\bibfnamefont {X.}~\bibnamefont
  {Chen}}, \bibinfo {author} {\bibfnamefont {Z.-C.}\ \bibnamefont {Gu}},\ and\
  \bibinfo {author} {\bibfnamefont {X.-G.}\ \bibnamefont {Wen}},\ }\bibfield
  {title} {\bibinfo {title} {Complete classification of one-dimensional gapped
  quantum phases in interacting spin systems},\ }\href
  {https://doi.org/10.1103/PhysRevB.84.235128} {\bibfield  {journal} {\bibinfo
  {journal} {Phys. Rev. B}\ }\textbf {\bibinfo {volume} {84}},\ \bibinfo
  {pages} {235128} (\bibinfo {year} {2011})}\BibitemShut {NoStop}%
\bibitem [{\citenamefont {Schuch}\ \emph {et~al.}(2011)\citenamefont {Schuch},
  \citenamefont {P\'erez-Garc\'{\i}a},\ and\ \citenamefont {Cirac}}]{Class4}%
  \BibitemOpen
  \bibfield  {author} {\bibinfo {author} {\bibfnamefont {N.}~\bibnamefont
  {Schuch}}, \bibinfo {author} {\bibfnamefont {D.}~\bibnamefont
  {P\'erez-Garc\'{\i}a}},\ and\ \bibinfo {author} {\bibfnamefont
  {I.}~\bibnamefont {Cirac}},\ }\bibfield  {title} {\bibinfo {title}
  {Classifying quantum phases using matrix product states and projected
  entangled pair states},\ }\href {https://doi.org/10.1103/PhysRevB.84.165139}
  {\bibfield  {journal} {\bibinfo  {journal} {Phys. Rev. B}\ }\textbf {\bibinfo
  {volume} {84}},\ \bibinfo {pages} {165139} (\bibinfo {year}
  {2011})}\BibitemShut {NoStop}%
\bibitem [{\citenamefont {Santos}(2020)}]{Hierarchy}%
  \BibitemOpen
  \bibfield  {author} {\bibinfo {author} {\bibfnamefont {L.~H.}\ \bibnamefont
  {Santos}},\ }\bibfield  {title} {\bibinfo {title} {Parafermions in
  hierarchical fractional quantum {H}all states},\ }\href
  {https://doi.org/10.1103/PhysRevResearch.2.013232} {\bibfield  {journal}
  {\bibinfo  {journal} {Phys. Rev. Research}\ }\textbf {\bibinfo {volume}
  {2}},\ \bibinfo {pages} {013232} (\bibinfo {year} {2020})}\BibitemShut
  {NoStop}%
\bibitem [{\citenamefont {Barkeshli}\ and\ \citenamefont
  {Qi}(2014)}]{Bilayer1}%
  \BibitemOpen
  \bibfield  {author} {\bibinfo {author} {\bibfnamefont {M.}~\bibnamefont
  {Barkeshli}}\ and\ \bibinfo {author} {\bibfnamefont {X.-L.}\ \bibnamefont
  {Qi}},\ }\bibfield  {title} {\bibinfo {title} {Synthetic topological qubits
  in conventional bilayer quantum {H}all systems},\ }\href
  {https://doi.org/10.1103/PhysRevX.4.041035} {\bibfield  {journal} {\bibinfo
  {journal} {Phys. Rev. X}\ }\textbf {\bibinfo {volume} {4}},\ \bibinfo {pages}
  {041035} (\bibinfo {year} {2014})}\BibitemShut {NoStop}%
\bibitem [{\citenamefont {Barkeshli}\ \emph {et~al.}(2013)\citenamefont
  {Barkeshli}, \citenamefont {Jian},\ and\ \citenamefont {Qi}}]{Bilayer2}%
  \BibitemOpen
  \bibfield  {author} {\bibinfo {author} {\bibfnamefont {M.}~\bibnamefont
  {Barkeshli}}, \bibinfo {author} {\bibfnamefont {C.-M.}\ \bibnamefont
  {Jian}},\ and\ \bibinfo {author} {\bibfnamefont {X.-L.}\ \bibnamefont {Qi}},\
  }\bibfield  {title} {\bibinfo {title} {Twist defects and projective
  non-{A}belian braiding statistics},\ }\href
  {https://doi.org/10.1103/PhysRevB.87.045130} {\bibfield  {journal} {\bibinfo
  {journal} {Phys. Rev. B}\ }\textbf {\bibinfo {volume} {87}},\ \bibinfo
  {pages} {045130} (\bibinfo {year} {2013})}\BibitemShut {NoStop}%
\bibitem [{\citenamefont {Barkeshli}\ \emph {et~al.}(2014)\citenamefont
  {Barkeshli}, \citenamefont {Oreg},\ and\ \citenamefont {Qi}}]{App5}%
  \BibitemOpen
  \bibfield  {author} {\bibinfo {author} {\bibfnamefont {M.}~\bibnamefont
  {Barkeshli}}, \bibinfo {author} {\bibfnamefont {Y.}~\bibnamefont {Oreg}},\
  and\ \bibinfo {author} {\bibfnamefont {X.-L.}\ \bibnamefont {Qi}},\
  }\bibfield  {title} {\bibinfo {title} {Experimental proposal to detect
  topological ground state degeneracy},\ }\href
  {https://arxiv.org/abs/1401.3750} {\bibfield  {journal} {\bibinfo  {journal}
  {arXiv:1401.3750}}}\BibitemShut {NoStop}%
\bibitem [{\citenamefont {Klinovaja}\ \emph {et~al.}(2014)\citenamefont
  {Klinovaja}, \citenamefont {Yacoby},\ and\ \citenamefont {Loss}}]{Loss2}%
  \BibitemOpen
  \bibfield  {author} {\bibinfo {author} {\bibfnamefont {J.}~\bibnamefont
  {Klinovaja}}, \bibinfo {author} {\bibfnamefont {A.}~\bibnamefont {Yacoby}},\
  and\ \bibinfo {author} {\bibfnamefont {D.}~\bibnamefont {Loss}},\ }\bibfield
  {title} {\bibinfo {title} {{K}ramers pairs of {M}ajorana fermions and
  parafermions in fractional topological insulators},\ }\href
  {https://doi.org/10.1103/PhysRevB.90.155447} {\bibfield  {journal} {\bibinfo
  {journal} {Phys. Rev. B}\ }\textbf {\bibinfo {volume} {90}},\ \bibinfo
  {pages} {155447} (\bibinfo {year} {2014})}\BibitemShut {NoStop}%
\bibitem [{\citenamefont {Orth}\ \emph {et~al.}(2015)\citenamefont {Orth},
  \citenamefont {Tiwari}, \citenamefont {Meng},\ and\ \citenamefont
  {Schmidt}}]{Tiwari15}%
  \BibitemOpen
  \bibfield  {author} {\bibinfo {author} {\bibfnamefont {C.~P.}\ \bibnamefont
  {Orth}}, \bibinfo {author} {\bibfnamefont {R.~P.}\ \bibnamefont {Tiwari}},
  \bibinfo {author} {\bibfnamefont {T.}~\bibnamefont {Meng}},\ and\ \bibinfo
  {author} {\bibfnamefont {T.~L.}\ \bibnamefont {Schmidt}},\ }\bibfield
  {title} {\bibinfo {title} {Non-{A}belian parafermions in
  time-reversal-invariant interacting helical systems},\ }\href
  {https://doi.org/10.1103/PhysRevB.91.081406} {\bibfield  {journal} {\bibinfo
  {journal} {Phys. Rev. B}\ }\textbf {\bibinfo {volume} {91}},\ \bibinfo
  {pages} {081406} (\bibinfo {year} {2015})}\BibitemShut {NoStop}%
\bibitem [{\citenamefont {Klinovaja}\ and\ \citenamefont
  {Loss}(2015)}]{Loss15}%
  \BibitemOpen
  \bibfield  {author} {\bibinfo {author} {\bibfnamefont {J.}~\bibnamefont
  {Klinovaja}}\ and\ \bibinfo {author} {\bibfnamefont {D.}~\bibnamefont
  {Loss}},\ }\bibfield  {title} {\bibinfo {title} {Fractional charge and spin
  states in topological insulator constrictions},\ }\href
  {https://doi.org/10.1103/PhysRevB.92.121410} {\bibfield  {journal} {\bibinfo
  {journal} {Phys. Rev. B}\ }\textbf {\bibinfo {volume} {92}},\ \bibinfo
  {pages} {121410} (\bibinfo {year} {2015})}\BibitemShut {NoStop}%
\bibitem [{\citenamefont {Peng}\ \emph {et~al.}(2016)\citenamefont {Peng},
  \citenamefont {Vinkler-Aviv}, \citenamefont {Brouwer}, \citenamefont
  {Glazman},\ and\ \citenamefont {von Oppen}}]{GlazmanPRL}%
  \BibitemOpen
  \bibfield  {author} {\bibinfo {author} {\bibfnamefont {Y.}~\bibnamefont
  {Peng}}, \bibinfo {author} {\bibfnamefont {Y.}~\bibnamefont {Vinkler-Aviv}},
  \bibinfo {author} {\bibfnamefont {P.~W.}\ \bibnamefont {Brouwer}}, \bibinfo
  {author} {\bibfnamefont {L.~I.}\ \bibnamefont {Glazman}},\ and\ \bibinfo
  {author} {\bibfnamefont {F.}~\bibnamefont {von Oppen}},\ }\bibfield  {title}
  {\bibinfo {title} {Parity anomaly and spin transmutation in quantum spin hall
  {J}osephson junctions},\ }\href
  {https://doi.org/10.1103/PhysRevLett.117.267001} {\bibfield  {journal}
  {\bibinfo  {journal} {Phys. Rev. Lett.}\ }\textbf {\bibinfo {volume} {117}},\
  \bibinfo {pages} {267001} (\bibinfo {year} {2016})}\BibitemShut {NoStop}%
\bibitem [{\citenamefont {Laubscher}\ \emph {et~al.}(2019)\citenamefont
  {Laubscher}, \citenamefont {Loss},\ and\ \citenamefont
  {Klinovaja}}]{Jelena19}%
  \BibitemOpen
  \bibfield  {author} {\bibinfo {author} {\bibfnamefont {K.}~\bibnamefont
  {Laubscher}}, \bibinfo {author} {\bibfnamefont {D.}~\bibnamefont {Loss}},\
  and\ \bibinfo {author} {\bibfnamefont {J.}~\bibnamefont {Klinovaja}},\
  }\bibfield  {title} {\bibinfo {title} {Fractional topological
  superconductivity and parafermion corner states},\ }\href
  {https://doi.org/10.1103/PhysRevResearch.1.032017} {\bibfield  {journal}
  {\bibinfo  {journal} {Phys. Rev. Research}\ }\textbf {\bibinfo {volume}
  {1}},\ \bibinfo {pages} {032017} (\bibinfo {year} {2019})}\BibitemShut
  {NoStop}%
\bibitem [{\citenamefont {Fleckenstein}\ \emph {et~al.}(2019)\citenamefont
  {Fleckenstein}, \citenamefont {Ziani},\ and\ \citenamefont
  {Trauzettel}}]{Trauzettel19}%
  \BibitemOpen
  \bibfield  {author} {\bibinfo {author} {\bibfnamefont {C.}~\bibnamefont
  {Fleckenstein}}, \bibinfo {author} {\bibfnamefont {N.~T.}\ \bibnamefont
  {Ziani}},\ and\ \bibinfo {author} {\bibfnamefont {B.}~\bibnamefont
  {Trauzettel}},\ }\bibfield  {title} {\bibinfo {title} {$\mathbb{Z}_{4}$
  parafermions in weakly interacting superconducting constrictions at the
  helical edge of quantum spin {H}all insulators},\ }\href
  {https://doi.org/10.1103/PhysRevLett.122.066801} {\bibfield  {journal}
  {\bibinfo  {journal} {Phys. Rev. Lett.}\ }\textbf {\bibinfo {volume} {122}},\
  \bibinfo {pages} {066801} (\bibinfo {year} {2019})}\BibitemShut {NoStop}%
\bibitem [{\citenamefont {Fendley}(2012)}]{Fendley12}%
  \BibitemOpen
  \bibfield  {author} {\bibinfo {author} {\bibfnamefont {P.}~\bibnamefont
  {Fendley}},\ }\bibfield  {title} {\bibinfo {title} {Parafermionic edge zero
  modes in $\mathbb{Z}_{n}$-invariant spin chains},\ }\href
  {https://doi.org/10.1088/1742-5468/2012/11/P11020} {\bibfield  {journal}
  {\bibinfo  {journal} {Journal of Statistical Mechanics: Theory and
  Experiment}\ }\textbf {\bibinfo {volume} {2012}},\ \bibinfo {pages} {P11020}
  (\bibinfo {year} {2012})}\BibitemShut {NoStop}%
\bibitem [{\citenamefont {Oreg}\ \emph {et~al.}(2014)\citenamefont {Oreg},
  \citenamefont {Sela},\ and\ \citenamefont {Stern}}]{Sela14}%
  \BibitemOpen
  \bibfield  {author} {\bibinfo {author} {\bibfnamefont {Y.}~\bibnamefont
  {Oreg}}, \bibinfo {author} {\bibfnamefont {E.}~\bibnamefont {Sela}},\ and\
  \bibinfo {author} {\bibfnamefont {A.}~\bibnamefont {Stern}},\ }\bibfield
  {title} {\bibinfo {title} {Fractional helical liquids in quantum wires},\
  }\href {https://doi.org/10.1103/PhysRevB.89.115402} {\bibfield  {journal}
  {\bibinfo  {journal} {Phys. Rev. B}\ }\textbf {\bibinfo {volume} {89}},\
  \bibinfo {pages} {115402} (\bibinfo {year} {2014})}\BibitemShut {NoStop}%
\bibitem [{\citenamefont {Klinovaja}\ and\ \citenamefont
  {Loss}(2014{\natexlab{a}})}]{Loss1}%
  \BibitemOpen
  \bibfield  {author} {\bibinfo {author} {\bibfnamefont {J.}~\bibnamefont
  {Klinovaja}}\ and\ \bibinfo {author} {\bibfnamefont {D.}~\bibnamefont
  {Loss}},\ }\bibfield  {title} {\bibinfo {title} {Parafermions in an
  interacting nanowire bundle},\ }\href
  {https://doi.org/10.1103/PhysRevLett.112.246403} {\bibfield  {journal}
  {\bibinfo  {journal} {Phys. Rev. Lett.}\ }\textbf {\bibinfo {volume} {112}},\
  \bibinfo {pages} {246403} (\bibinfo {year} {2014}{\natexlab{a}})}\BibitemShut
  {NoStop}%
\bibitem [{\citenamefont {Klinovaja}\ and\ \citenamefont
  {Loss}(2014{\natexlab{b}})}]{Loss14}%
  \BibitemOpen
  \bibfield  {author} {\bibinfo {author} {\bibfnamefont {J.}~\bibnamefont
  {Klinovaja}}\ and\ \bibinfo {author} {\bibfnamefont {D.}~\bibnamefont
  {Loss}},\ }\bibfield  {title} {\bibinfo {title} {Time-reversal invariant
  parafermions in interacting {R}ashba nanowires},\ }\href
  {https://doi.org/10.1103/PhysRevB.90.045118} {\bibfield  {journal} {\bibinfo
  {journal} {Phys. Rev. B}\ }\textbf {\bibinfo {volume} {90}},\ \bibinfo
  {pages} {045118} (\bibinfo {year} {2014}{\natexlab{b}})}\BibitemShut
  {NoStop}%
\bibitem [{\citenamefont {Thakurathi}\ \emph {et~al.}(2017)\citenamefont
  {Thakurathi}, \citenamefont {Loss},\ and\ \citenamefont
  {Klinovaja}}]{Manisha17}%
  \BibitemOpen
  \bibfield  {author} {\bibinfo {author} {\bibfnamefont {M.}~\bibnamefont
  {Thakurathi}}, \bibinfo {author} {\bibfnamefont {D.}~\bibnamefont {Loss}},\
  and\ \bibinfo {author} {\bibfnamefont {J.}~\bibnamefont {Klinovaja}},\
  }\bibfield  {title} {\bibinfo {title} {Floquet majorana fermions and
  parafermions in driven {R}ashba nanowires},\ }\href
  {https://doi.org/10.1103/PhysRevB.95.155407} {\bibfield  {journal} {\bibinfo
  {journal} {Phys. Rev. B}\ }\textbf {\bibinfo {volume} {95}},\ \bibinfo
  {pages} {155407} (\bibinfo {year} {2017})}\BibitemShut {NoStop}%
\bibitem [{\citenamefont {Chew}\ \emph {et~al.}(2018)\citenamefont {Chew},
  \citenamefont {Mross},\ and\ \citenamefont {Alicea}}]{Mross18}%
  \BibitemOpen
  \bibfield  {author} {\bibinfo {author} {\bibfnamefont {A.}~\bibnamefont
  {Chew}}, \bibinfo {author} {\bibfnamefont {D.~F.}\ \bibnamefont {Mross}},\
  and\ \bibinfo {author} {\bibfnamefont {J.}~\bibnamefont {Alicea}},\
  }\bibfield  {title} {\bibinfo {title} {Fermionized parafermions and
  symmetry-enriched {M}ajorana modes},\ }\href
  {https://doi.org/10.1103/PhysRevB.98.085143} {\bibfield  {journal} {\bibinfo
  {journal} {Phys. Rev. B}\ }\textbf {\bibinfo {volume} {98}},\ \bibinfo
  {pages} {085143} (\bibinfo {year} {2018})}\BibitemShut {NoStop}%
\bibitem [{\citenamefont {Mazza}\ \emph {et~al.}(2018)\citenamefont {Mazza},
  \citenamefont {Iemini}, \citenamefont {Dalmonte},\ and\ \citenamefont
  {Mora}}]{Mora18}%
  \BibitemOpen
  \bibfield  {author} {\bibinfo {author} {\bibfnamefont {L.}~\bibnamefont
  {Mazza}}, \bibinfo {author} {\bibfnamefont {F.}~\bibnamefont {Iemini}},
  \bibinfo {author} {\bibfnamefont {M.}~\bibnamefont {Dalmonte}},\ and\
  \bibinfo {author} {\bibfnamefont {C.}~\bibnamefont {Mora}},\ }\bibfield
  {title} {\bibinfo {title} {Nontopological parafermions in a one-dimensional
  fermionic model with even multiplet pairing},\ }\href
  {https://doi.org/10.1103/PhysRevB.98.201109} {\bibfield  {journal} {\bibinfo
  {journal} {Phys. Rev. B}\ }\textbf {\bibinfo {volume} {98}},\ \bibinfo
  {pages} {201109} (\bibinfo {year} {2018})}\BibitemShut {NoStop}%
\bibitem [{\citenamefont {Ronetti}\ \emph {et~al.}(2021)\citenamefont
  {Ronetti}, \citenamefont {Loss},\ and\ \citenamefont
  {Klinovaja}}]{Ronetti21}%
  \BibitemOpen
  \bibfield  {author} {\bibinfo {author} {\bibfnamefont {F.}~\bibnamefont
  {Ronetti}}, \bibinfo {author} {\bibfnamefont {D.}~\bibnamefont {Loss}},\ and\
  \bibinfo {author} {\bibfnamefont {J.}~\bibnamefont {Klinovaja}},\ }\bibfield
  {title} {\bibinfo {title} {Clock model and parafermions in {R}ashba
  nanowires},\ }\href {https://doi.org/10.1103/PhysRevB.103.235410} {\bibfield
  {journal} {\bibinfo  {journal} {Phys. Rev. B}\ }\textbf {\bibinfo {volume}
  {103}},\ \bibinfo {pages} {235410} (\bibinfo {year} {2021})}\BibitemShut
  {NoStop}%
\bibitem [{\citenamefont {Snizhko}\ \emph
  {et~al.}(2018{\natexlab{b}})\citenamefont {Snizhko}, \citenamefont {Egger},\
  and\ \citenamefont {Gefen}}]{Kyrylo1}%
  \BibitemOpen
  \bibfield  {author} {\bibinfo {author} {\bibfnamefont {K.}~\bibnamefont
  {Snizhko}}, \bibinfo {author} {\bibfnamefont {R.}~\bibnamefont {Egger}},\
  and\ \bibinfo {author} {\bibfnamefont {Y.}~\bibnamefont {Gefen}},\ }\bibfield
   {title} {\bibinfo {title} {Measurement and control of a {C}oulomb-blockaded
  parafermion box},\ }\href {https://doi.org/10.1103/PhysRevB.97.081405}
  {\bibfield  {journal} {\bibinfo  {journal} {Phys. Rev. B}\ }\textbf {\bibinfo
  {volume} {97}},\ \bibinfo {pages} {081405} (\bibinfo {year}
  {2018}{\natexlab{b}})}\BibitemShut {NoStop}%
\bibitem [{\citenamefont {Nielsen}\ \emph {et~al.}(2021)\citenamefont
  {Nielsen}, \citenamefont {Flensberg}, \citenamefont {Egger},\ and\ \citenamefont {Burrello}}]{App13}%
  \BibitemOpen
  \bibfield  {author} {\bibinfo {author} {\bibfnamefont {I. E.}~\bibnamefont
  {Nielsen}}, \bibinfo {author} {\bibfnamefont {K.}~\bibnamefont {Flensberg}}, 
  \bibinfo {author} {\bibfnamefont {R.}~\bibnamefont {Egger}},\
  and\ \bibinfo {author} {\bibfnamefont {M.}\ \bibnamefont {Burrello}},\
  }\bibfield  {title} {\bibinfo {title} {Readout of parafermionic states by transport measurements},\ }\href
  {https://arxiv.org/abs/2109.02300} {\bibfield  {journal} {\bibinfo  {journal}
  {arXiv:2109.02300}}}\BibitemShut {NoStop}%
\bibitem [{\citenamefont {Trebst}\ \emph {et~al.}(2008)\citenamefont {Trebst},
  \citenamefont {Troyer}, \citenamefont {Wang},\ and\ \citenamefont
  {Ludwig}}]{Fibonacci1}%
  \BibitemOpen
  \bibfield  {author} {\bibinfo {author} {\bibfnamefont {S.}~\bibnamefont
  {Trebst}}, \bibinfo {author} {\bibfnamefont {M.}~\bibnamefont {Troyer}},
  \bibinfo {author} {\bibfnamefont {Z.}~\bibnamefont {Wang}},\ and\ \bibinfo
  {author} {\bibfnamefont {A.~W.~W.}\ \bibnamefont {Ludwig}},\ }\bibfield
  {title} {\bibinfo {title} {A short introduction to {F}ibonacci anyon
  models},\ }\href {https://doi.org/10.1143/PTPS.176.384} {\bibfield  {journal}
  {\bibinfo  {journal} {Progress of Theoretical Physics Supplement}\ }\textbf
  {\bibinfo {volume} {176}},\ \bibinfo {pages} {384} (\bibinfo {year}
  {2008})}\BibitemShut {NoStop}%
\bibitem [{\citenamefont {Stoudenmire}\ \emph {et~al.}(2015)\citenamefont
  {Stoudenmire}, \citenamefont {Clarke}, \citenamefont {Mong},\ and\
  \citenamefont {Alicea}}]{Fibonacci3}%
  \BibitemOpen
  \bibfield  {author} {\bibinfo {author} {\bibfnamefont {E.~M.}\ \bibnamefont
  {Stoudenmire}}, \bibinfo {author} {\bibfnamefont {D.~J.}\ \bibnamefont
  {Clarke}}, \bibinfo {author} {\bibfnamefont {R.~S.~K.}\ \bibnamefont
  {Mong}},\ and\ \bibinfo {author} {\bibfnamefont {J.}~\bibnamefont {Alicea}},\
  }\bibfield  {title} {\bibinfo {title} {Assembling {F}ibonacci anyons from a
  $\mathbb{Z}_{3}$ parafermion lattice model},\ }\href
  {https://doi.org/10.1103/PhysRevB.91.235112} {\bibfield  {journal} {\bibinfo
  {journal} {Phys. Rev. B}\ }\textbf {\bibinfo {volume} {91}},\ \bibinfo
  {pages} {235112} (\bibinfo {year} {2015})}\BibitemShut {NoStop}%
\bibitem [{\citenamefont {Alicea}\ \emph {et~al.}(2011)\citenamefont {Alicea},
  \citenamefont {Oreg}, \citenamefont {Refael}, \citenamefont {von Oppen},\
  and\ \citenamefont {Fisher}}]{TJ0}%
  \BibitemOpen
  \bibfield  {author} {\bibinfo {author} {\bibfnamefont {J.}~\bibnamefont
  {Alicea}}, \bibinfo {author} {\bibfnamefont {Y.}~\bibnamefont {Oreg}},
  \bibinfo {author} {\bibfnamefont {G.}~\bibnamefont {Refael}}, \bibinfo
  {author} {\bibfnamefont {F.}~\bibnamefont {von Oppen}},\ and\ \bibinfo
  {author} {\bibfnamefont {M.~P.~A.}\ \bibnamefont {Fisher}},\ }\bibfield
  {title} {\bibinfo {title} {Non-{A}belian statistics and topological quantum
  information processing in 1d wire networks},\ }\href
  {https://doi.org/10.1038/nphys1915} {\bibfield  {journal} {\bibinfo
  {journal} {Nature Physics}\ }\textbf {\bibinfo {volume} {7}},\ \bibinfo
  {pages} {412} (\bibinfo {year} {2011})}\BibitemShut {NoStop}%
\bibitem [{\citenamefont {van Heck}\ \emph {et~al.}(2012)\citenamefont {van
  Heck}, \citenamefont {Akhmerov}, \citenamefont {Hassler}, \citenamefont
  {Burrello},\ and\ \citenamefont {Beenakker}}]{Beenakker1}%
  \BibitemOpen
  \bibfield  {author} {\bibinfo {author} {\bibfnamefont {B.}~\bibnamefont {van
  Heck}}, \bibinfo {author} {\bibfnamefont {A.~R.}\ \bibnamefont {Akhmerov}},
  \bibinfo {author} {\bibfnamefont {F.}~\bibnamefont {Hassler}}, \bibinfo
  {author} {\bibfnamefont {M.}~\bibnamefont {Burrello}},\ and\ \bibinfo
  {author} {\bibfnamefont {C.~W.~J.}\ \bibnamefont {Beenakker}},\ }\bibfield
  {title} {\bibinfo {title} {{C}oulomb-assisted braiding of {M}ajorana fermions
  in a {J}osephson junction array},\ }\href
  {https://doi.org/10.1088/1367-2630/14/3/035019} {\bibfield  {journal}
  {\bibinfo  {journal} {New J. Phys.}\ }\textbf {\bibinfo {volume} {14}},\
  \bibinfo {pages} {035019} (\bibinfo {year} {2012})}\BibitemShut {NoStop}%
\bibitem [{\citenamefont {Li}\ \emph {et~al.}(2016)\citenamefont {Li},
  \citenamefont {Neupert}, \citenamefont {Bernevig},\ and\ \citenamefont
  {Yazdani}}]{TJ3}%
  \BibitemOpen
  \bibfield  {author} {\bibinfo {author} {\bibfnamefont {J.}~\bibnamefont
  {Li}}, \bibinfo {author} {\bibfnamefont {T.}~\bibnamefont {Neupert}},
  \bibinfo {author} {\bibfnamefont {B.~A.}\ \bibnamefont {Bernevig}},\ and\
  \bibinfo {author} {\bibfnamefont {A.}~\bibnamefont {Yazdani}},\ }\bibfield
  {title} {\bibinfo {title} {Manipulating {M}ajorana zero modes on atomic rings
  with an external magnetic field},\ }\href
  {https://doi.org/10.1038/ncomms10395} {\bibfield  {journal} {\bibinfo
  {journal} {Nat. Commun.}\ }\textbf {\bibinfo {volume} {7}},\ \bibinfo {pages}
  {10395} (\bibinfo {year} {2016})}\BibitemShut {NoStop}%
\bibitem [{\citenamefont {Trif}\ and\ \citenamefont {Simon}(2019)}]{TJ1}%
  \BibitemOpen
  \bibfield  {author} {\bibinfo {author} {\bibfnamefont {M.}~\bibnamefont
  {Trif}}\ and\ \bibinfo {author} {\bibfnamefont {P.}~\bibnamefont {Simon}},\
  }\bibfield  {title} {\bibinfo {title} {Braiding of {M}ajorana fermions in a
  cavity},\ }\href {https://doi.org/10.1103/PhysRevLett.122.236803} {\bibfield
  {journal} {\bibinfo  {journal} {Phys. Rev. Lett.}\ }\textbf {\bibinfo
  {volume} {122}},\ \bibinfo {pages} {236803} (\bibinfo {year}
  {2019})}\BibitemShut {NoStop}%
\bibitem [{\citenamefont {Alvarado}\ \emph {et~al.}(2020)\citenamefont
  {Alvarado}, \citenamefont {Iks}, \citenamefont {Zazunov}, \citenamefont
  {Egger},\ and\ \citenamefont {Yeyati}}]{TJ2}%
  \BibitemOpen
  \bibfield  {author} {\bibinfo {author} {\bibfnamefont {M.}~\bibnamefont
  {Alvarado}}, \bibinfo {author} {\bibfnamefont {A.}~\bibnamefont {Iks}},
  \bibinfo {author} {\bibfnamefont {A.}~\bibnamefont {Zazunov}}, \bibinfo
  {author} {\bibfnamefont {R.}~\bibnamefont {Egger}},\ and\ \bibinfo {author}
  {\bibfnamefont {A.~L.}\ \bibnamefont {Yeyati}},\ }\bibfield  {title}
  {\bibinfo {title} {Boundary {G}reen's function approach for spinful
  single-channel and multichannel {M}ajorana nanowires},\ }\href
  {https://doi.org/10.1103/PhysRevB.101.094511} {\bibfield  {journal} {\bibinfo
   {journal} {Phys. Rev. B}\ }\textbf {\bibinfo {volume} {101}},\ \bibinfo
  {pages} {094511} (\bibinfo {year} {2020})}\BibitemShut {NoStop}%
\bibitem [{\citenamefont {Deb}\ \emph {et~al.}(2018)\citenamefont {Deb},
  \citenamefont {Sengupta},\ and\ \citenamefont {Sen}}]{TJ4}%
  \BibitemOpen
  \bibfield  {author} {\bibinfo {author} {\bibfnamefont {O.}~\bibnamefont
  {Deb}}, \bibinfo {author} {\bibfnamefont {K.}~\bibnamefont {Sengupta}},\ and\
  \bibinfo {author} {\bibfnamefont {D.}~\bibnamefont {Sen}},\ }\bibfield
  {title} {\bibinfo {title} {{J}osephson junctions of multiple superconducting
  wires},\ }\href {https://doi.org/10.1103/PhysRevB.97.174518} {\bibfield
  {journal} {\bibinfo  {journal} {Phys. Rev. B}\ }\textbf {\bibinfo {volume}
  {97}},\ \bibinfo {pages} {174518} (\bibinfo {year} {2018})}\BibitemShut
  {NoStop}%
\bibitem [{\citenamefont {Peralta~Gavensky}\ \emph {et~al.}(2019)\citenamefont
  {Peralta~Gavensky}, \citenamefont {Usaj},\ and\ \citenamefont
  {Balseiro}}]{TJ5}%
  \BibitemOpen
  \bibfield  {author} {\bibinfo {author} {\bibfnamefont {L.}~\bibnamefont
  {Peralta~Gavensky}}, \bibinfo {author} {\bibfnamefont {G.}~\bibnamefont
  {Usaj}},\ and\ \bibinfo {author} {\bibfnamefont {C.~A.}\ \bibnamefont
  {Balseiro}},\ }\bibfield  {title} {\bibinfo {title} {Topological phase
  diagram of a three-terminal {J}osephson junction: From the conventional to
  the {M}ajorana regime},\ }\href {https://doi.org/10.1103/PhysRevB.100.014514}
  {\bibfield  {journal} {\bibinfo  {journal} {Phys. Rev. B}\ }\textbf {\bibinfo
  {volume} {100}},\ \bibinfo {pages} {014514} (\bibinfo {year}
  {2019})}\BibitemShut {NoStop}%
\bibitem [{\citenamefont {Meyer}\ and\ \citenamefont {Houzet}(2021)}]{TJ6}%
  \BibitemOpen
  \bibfield  {author} {\bibinfo {author} {\bibfnamefont {J.~S.}\ \bibnamefont
  {Meyer}}\ and\ \bibinfo {author} {\bibfnamefont {M.}~\bibnamefont {Houzet}},\
  }\bibfield  {title} {\bibinfo {title} {Conductance quantization in
  topological {J}osephson trijunctions},\ }\href
  {https://doi.org/10.1103/PhysRevB.103.174504} {\bibfield  {journal} {\bibinfo
   {journal} {Phys. Rev. B}\ }\textbf {\bibinfo {volume} {103}},\ \bibinfo
  {pages} {174504} (\bibinfo {year} {2021})}\BibitemShut {NoStop}%
\bibitem [{\citenamefont {Jonckheere}\ \emph {et~al.}(2019)\citenamefont
  {Jonckheere}, \citenamefont {Rech}, \citenamefont {Zazunov}, \citenamefont
  {Egger}, \citenamefont {Yeyati},\ and\ \citenamefont {Martin}}]{TJ7}%
  \BibitemOpen
  \bibfield  {author} {\bibinfo {author} {\bibfnamefont {T.}~\bibnamefont
  {Jonckheere}}, \bibinfo {author} {\bibfnamefont {J.}~\bibnamefont {Rech}},
  \bibinfo {author} {\bibfnamefont {A.}~\bibnamefont {Zazunov}}, \bibinfo
  {author} {\bibfnamefont {R.}~\bibnamefont {Egger}}, \bibinfo {author}
  {\bibfnamefont {A.~L.}\ \bibnamefont {Yeyati}},\ and\ \bibinfo {author}
  {\bibfnamefont {T.}~\bibnamefont {Martin}},\ }\bibfield  {title} {\bibinfo
  {title} {Giant shot noise from {M}ajorana zero modes in topological
  trijunctions},\ }\href {https://doi.org/10.1103/PhysRevLett.122.097003}
  {\bibfield  {journal} {\bibinfo  {journal} {Phys. Rev. Lett.}\ }\textbf
  {\bibinfo {volume} {122}},\ \bibinfo {pages} {097003} (\bibinfo {year}
  {2019})}\BibitemShut {NoStop}%
\bibitem [{\citenamefont {Wen}(1990)}]{Wen1990}%
  \BibitemOpen
  \bibfield  {author} {\bibinfo {author} {\bibfnamefont {X.~G.}\ \bibnamefont
  {Wen}},\ }\bibfield  {title} {\bibinfo {title} {Electrodynamical properties
  of gapless edge excitations in the fractional quantum {H}all states},\ }\href
  {https://doi.org/10.1103/PhysRevLett.64.2206} {\bibfield  {journal} {\bibinfo
   {journal} {Phys. Rev. Lett.}\ }\textbf {\bibinfo {volume} {64}},\ \bibinfo
  {pages} {2206} (\bibinfo {year} {1990})}\BibitemShut {NoStop}%
\bibitem [{Foo()}]{Foot1}%
  \BibitemOpen
  \href@noop {} {}\bibinfo {note} {Our results are also valid (qualitatively)
  for a trijunction comprising the boundary of spin-unpolarized $\nu =
  \frac{2}{3}$ QH puddles. In this case, $s$-wave superconducting segments may
  be employed to gap out the counter-propagating edge modes.}\BibitemShut
  {Stop}%
\bibitem [{SM()}]{SM}%
  \BibitemOpen
  \href@noop {} {}\bibinfo {note} {See Supplemental Material for further
  details.}\BibitemShut {Stop}%
\bibitem [{\citenamefont {Chen}\ and\ \citenamefont
  {Burnell}(2016)}]{Burnell16}%
  \BibitemOpen
  \bibfield  {author} {\bibinfo {author} {\bibfnamefont {C.}~\bibnamefont
  {Chen}}\ and\ \bibinfo {author} {\bibfnamefont {F.~J.}\ \bibnamefont
  {Burnell}},\ }\bibfield  {title} {\bibinfo {title} {Tunable splitting of the
  ground-state degeneracy in quasi-one-dimensional parafermion systems},\
  }\href {https://doi.org/10.1103/PhysRevLett.116.106405} {\bibfield  {journal}
  {\bibinfo  {journal} {Phys. Rev. Lett.}\ }\textbf {\bibinfo {volume} {116}},\
  \bibinfo {pages} {106405} (\bibinfo {year} {2016})}\BibitemShut {NoStop}%
\bibitem [{\citenamefont {Wen}(1991)}]{Wen91PRB}%
  \BibitemOpen
  \bibfield  {author} {\bibinfo {author} {\bibfnamefont {X.-G.}\ \bibnamefont
  {Wen}},\ }\bibfield  {title} {\bibinfo {title} {Edge transport properties of
  the fractional quantum {H}all states and weak-impurity scattering of a
  one-dimensional charge-density wave},\ }\href
  {https://doi.org/10.1103/PhysRevB.44.5708} {\bibfield  {journal} {\bibinfo
  {journal} {Phys. Rev. B}\ }\textbf {\bibinfo {volume} {44}},\ \bibinfo
  {pages} {5708} (\bibinfo {year} {1991})}\BibitemShut {NoStop}%
\bibitem [{\citenamefont {Maciazek}\ and\ \citenamefont
  {Sawicki}(2019)}]{Tomasz0}%
  \BibitemOpen
  \bibfield  {author} {\bibinfo {author} {\bibfnamefont {T.}~\bibnamefont
  {Maciazek}}\ and\ \bibinfo {author} {\bibfnamefont {A.}~\bibnamefont
  {Sawicki}},\ }\bibfield  {title} {\bibinfo {title} {Non-{A}belian quantum
  statistics on graphs},\ }\href {https://doi.org/10.1007/s00220-019-03583-5}
  {\bibfield  {journal} {\bibinfo  {journal} {Commun. Math. Phys.}\ }\textbf
  {\bibinfo {volume} {371}},\ \bibinfo {pages} {921} (\bibinfo {year}
  {2019})}\BibitemShut {NoStop}%
\bibitem [{\citenamefont {Maciazek}\ and\ \citenamefont {An}(2020)}]{Tomasz1}%
  \BibitemOpen
  \bibfield  {author} {\bibinfo {author} {\bibfnamefont {T.}~\bibnamefont
  {Maciazek}}\ and\ \bibinfo {author} {\bibfnamefont {B.~H.}\ \bibnamefont
  {An}},\ }\bibfield  {title} {\bibinfo {title} {Universal properties of anyon
  braiding on one-dimensional wire networks},\ }\href
  {https://doi.org/10.1103/PhysRevB.102.201407} {\bibfield  {journal} {\bibinfo
   {journal} {Phys. Rev. B}\ }\textbf {\bibinfo {volume} {102}},\ \bibinfo
  {pages} {201407} (\bibinfo {year} {2020})}\BibitemShut {NoStop}%
\bibitem [{\citenamefont {An}\ and\ \citenamefont {Maciazek}(2021)}]{Tomasz2}%
  \BibitemOpen
  \bibfield  {author} {\bibinfo {author} {\bibfnamefont {B.~H.}\ \bibnamefont
  {An}}\ and\ \bibinfo {author} {\bibfnamefont {T.}~\bibnamefont {Maciazek}},\
  }\bibfield  {title} {\bibinfo {title} {Geometric presentations of braid
  groups for particles on a graph},\ }\href
  {https://doi.org/10.1007/s00220-021-04095-x} {\bibfield  {journal} {\bibinfo
  {journal} {Comm. Math. Phys.}\ }\textbf {\bibinfo {volume} {384}},\ \bibinfo
  {pages} {1109} (\bibinfo {year} {2021})}\BibitemShut {NoStop}%
\end{thebibliography}
\end{document}